\documentclass[prb,twocolumn,notitlepage,showpacs,amsmath,amstex,amssymb,citeautoscript,longbibliography]{revtex4-1}
\pdfoutput=1
\usepackage{natbib}
\usepackage[english]{babel}
\usepackage{letltxmacro}
\LetLtxMacro{\ORIGselectlanguage}{\selectlanguage}
\makeatletter
\DeclareRobustCommand{\selectlanguage}[1]{%
  \@ifundefined{alias@\string#1}
    {\ORIGselectlanguage{#1}}
    {\begingroup\edef\x{\endgroup
       \noexpand\ORIGselectlanguage{\@nameuse{alias@#1}}}\x}%
}
\newcommand{\definelanguagealias}[2]{%
  \@namedef{alias@#1}{#2}%
}
\makeatother
\definelanguagealias{en}{english}
\definelanguagealias{English}{english}

\usepackage{amsmath}
\usepackage{amsfonts}
\usepackage{amssymb}
\usepackage{mathtools}
\usepackage{dsfont}
\usepackage{braket}
\usepackage{soul}
\usepackage{blkarray}
\usepackage{multirow}

\usepackage{tikz}
\usetikzlibrary{positioning}

\usepackage{hyperref}
\hypersetup{
    bookmarks=false,         
    unicode=false,          
    pdftoolbar=false,        
    pdfmenubar=true,        
    pdffitwindow=false,     
    pdfstartview={FitH},    
    pdftitle={Quantum scarred eigenstates in a Rydberg atom chain: entanglement, breakdown of thermalization, and stability to perturbations},    
    pdfauthor={C. J. Turner, A. A. Michailidis, D. A. Abanin, M. Serbyn, and Z. Papic},     
    pdfsubject={},   
    pdfcreator={},   
    pdfproducer={}, 
    pdfkeywords={quantum scars} {non-ergodic dynamics}, 
    pdfnewwindow=true,      
    colorlinks=true,       
    linkcolor=black,          
    citecolor=blue,        
    filecolor=magenta,      
    urlcolor=blue           
}

\newcommand\ProvideMathOperator[2]{\ifdefined#1\else\DeclareMathOperator{#1}{#2}\fi}
\ProvideMathOperator{\tr}{tr}
\ProvideMathOperator{\var}{var}

\begin{document}
\title{Quantum scarred eigenstates in a Rydberg atom chain: entanglement, breakdown of thermalization, and stability to perturbations}
\author{C. J. Turner$^1$, A. A. Michailidis$^2$, D. A. Abanin$^3$, M. Serbyn$^2$, and Z. Papi\'c$^1$}
\affiliation{$^1$School of Physics and Astronomy, University of Leeds, Leeds LS2 9JT, United Kingdom}
\affiliation{$^2$IST Austria, Am Campus 1, 3400 Klosterneuburg, Austria}
\affiliation{$^3$Department of Theoretical Physics, University of Geneva, 24 quai Ernest-Ansermet, 1211 Geneva, Switzerland}
\date{\today}
\begin{abstract}

Recent realization of a kinetically-constrained chain of Rydberg atoms by Bernien~\emph{et al.}~[\href{https://doi.org/10.1038/nature24622}{Nature {\bf 551}, 579 (2017)}] resulted in the observation of unusual revivals in the many-body quantum dynamics.  In our previous work, Turner~\emph{et al.}~[\href{https://arxiv.org/pdf/1711.03528.pdf}{arXiv:1711.03528}], such dynamics was attributed to the existence of ``quantum scarred" eigenstates in the many-body   spectrum of the experimentally realized model. Here we present a detailed study of the eigenstate properties of the same model. We find that the majority of the eigenstates exhibit anomalous thermalization: the observable expectation values converge to their Gibbs ensemble values, but parametrically slower compared to the predictions of the eigenstate thermalization hypothesis~(ETH). Amidst the thermalizing spectrum, we identify non-ergodic eigenstates that strongly violate the ETH, whose number grows polynomially with system size. Previously, the same eigenstates were identified via large overlaps with certain product states, and were used to explain the revivals observed in experiment. Here we find that these eigenstates, in addition to highly atypical expectation values of local observables, also exhibit sub-thermal entanglement entropy that scales logarithmically with the system size.  Moreover, we identify an additional class of quantum scarred eigenstates, and discuss their manifestations in the dynamics starting from initial product states.  We use forward scattering approximation to describe the structure and physical properties of quantum-scarred eigenstates.  Finally, we discuss the stability of quantum scars to various perturbations.  We observe that quantum scars remain robust when the introduced perturbation is compatible with the forward scattering approximation. In contrast, the perturbations which most efficiently destroy quantum scars also lead to the restoration of ``canonical'' thermalization. 
\end{abstract}
\maketitle

\section{Introduction}

In recent years, significant efforts have been focused on understanding the process of \emph{quantum thermalization}, i.e., the approach to equilibrium of quantum systems which are well-isolated from any external thermal bath. The considerable interest in this problem has come hand in hand with the experimental advances in controllable, quantum-coherent systems of ultracold atoms~\cite{Kinoshita06, Bloch15}, trapped ions~\cite{Monroe16}, and nitrogen-vacancy spins in diamond~\cite{Lukin16}. These systems allow  one to realize highly tunable lattice models of interacting spins, bosons or fermions, and to characterize their quantum thermalization~\cite{Kaufman2016}.

The process of quantum thermalization is believed to be controlled by the properties of the system's many-body eigenstates, in which physical observables have thermal expectation values. This scenario where each of the system's eigenstates forms its own ``thermal ensemble" is known as the {\it Eigenstate Thermalization Hypothesis} (ETH)~\cite{DeutschETH,SrednickiETH}. Despite the lack of a formal proof of the ETH, various numerical studies of the systems of spins, fermions, and bosons in 1D and 2D~\cite{Rigol07,RigolNature} suggest that in many cases when the system thermalizes, \emph{all} of its highly excited eigenstates obey the ETH~\cite{Huse14}, i.e., they are typical thermal states and akin to random vectors.

However, not all quantum systems obey the ETH. Indeed, in integrable systems~\cite{Sutherland} and many-body localized phases~\cite{Basko06,Serbyn13-1,Huse13} the ETH is strongly violated due to the appearance of extensively many conserved operators, $K_i$, which commute with the system's Hamiltonian, $[ H, K_i] = 0$. For example, in many-body localized phases $K_i$ correspond to deformations of simple number operators of Anderson localized single-particle orbitals~\cite{Anderson58}. The presence of such operators prevents the system, initialized in a random state, from fully exploring all allowed configurations in the Hilbert space, leading to strong ergodicity breaking. 

Despite significant progress in theoretical understanding of fully thermalizing~\cite{Alessiorev} and many-body localized systems,~\cite{AbaninRev} much less is known about the possibility of more subtle intermediate behaviors. In particular, can ergodicity be broken in interacting, translationally invariant quantum systems? In the classical case, non-thermalizing behavior without disorder is well-known in the context of structural glasses~\cite{Binder2011,Berthier2011,Biroli2013}. The mechanism of this type of behavior is the excluded volume interactions that impose kinetic constraints on the dynamics~\cite{Fredrickson1984,Palmer1984}.  Similar type of physics has recently been explored in quantum systems where a ``quasi many-body localized" behavior was proposed to occur in the absence of disorder~\cite{carleo2012localization, Huveneers13, Muller, Yao14, Papic, Juan15, QDLEssler, Kim2016,Yarloo2017,michailidis2017slow,Lan2017_2,Smith2017,brenes2017many}. 

Recently, a striking phenomenon suggestive of a different mechanism of weak ergodicity breaking was discovered experimentally~\cite{Bernien2017}. A Rydberg atoms platform~\cite{Schauss2012,Labuhn2016,Bernien2017} was used to realize a quantum model with kinetic constraints induced by strong nearest-neighbour repulsion between atoms in excited states. The experiment observed persistent many-body revivals after a quench from a N\'eel-type state. In contrast, other initial configurations probed in the experiment exhibited fast equilibration without any revivals. The unexpected many-body revivals are inconsistent with ergodicity and thermalization. Moreover, the strong dependence of relaxation dynamics on the initial state is unusual: for example, many-body localized systems fail to thermalize \emph{irrespective} of their initial state.~\cite{AbaninRev} 

In Ref.~\onlinecite{Turner2017}, we attributed the observed slow equilibration and revivals to a special band of highly non-thermal eigenstates, and proposed an analogy to \emph{quantum scars} first discovered in single-particle chaotic billiards~\cite{Heller84}. In the semiclassical quantization of single-particle chaotic billiards, scars represent an enhancement of eigenfunction density along the trajectories of classical periodic orbits. Even though such classical orbits are unstable, they nevertheless leave a ``scar" on the states of the corresponding quantum system.  The enhancement of the eigenstate probability density near a classical orbit implies the breakdown of ergodicity in scarred eigenstates. Moreover, quantum scars are surprisingly robust to perturbations, and their experimental signatures have been detected in a variety of systems, including microwave cavities,~\cite{Sridhar1991} quantum dots,~\cite{Marcus1992} and quantum wells.~\cite{Wilkinson1996}

In the single-particle case quantum scars are often probed by preparing a particle in a Gaussian wave packet  localized near the classical periodic trajectory. By analogy, in Ref.~\onlinecite{Turner2017} an anomalous concentration of special eigenstates in the Hilbert space was demonstrated, thus providing phenomenological support to the quantum scar analogy. In addition, we presented an explicit method that allowed us to construct these special eigenstates, and demonstrated the absence integrability in the studied model. At the same time, many properties of these quantum scarred eigenstates remained unexplored.  For example, what is the  entanglement structure of special eigenstates? How many different classes of quantum scars exist? What is the relation between the presence of quantum scarred eigenstates and thermalization?

In this paper we present a detailed study of the properties of scarred eigenstates which addresses the above questions. We start by introducing the model of the experimentally-realized Rydberg chain in Sec.~\ref{sec:model},  and discuss the structure of its Hilbert space.  In Sec.~\ref{sec:eth} we investigate properties of  the eigenstates of this model from the point of view of the ETH and quantum entanglement. We find that the many-body spectrum is distinguished by the presence of 
special eigenstates, which have atypical expectation values of local observables and thus strongly violate the ETH. At the same time, the majority of eigenstates in the spectrum appear thermal; yet, the diagonal  matrix elements of  local observables converge to the prediction of the Gibbs ensemble much more slowly compared to other, non-constrained models. We  further test the ETH by studying off-diagonal matrix elements of local operators, and find that the spectral function of local observables has an unusual, non-monotonic form, with a peak at the frequency coinciding with the energy separation between special eigenstates. We identify the anomalous expectation values of local observables, low entanglement entropy, and enhanced overlap with certain product states to be the key features distinguishing quantum scarred eigenstates. Using these signatures, we find an additional family of quantum scarred eigenstates. These states manifest themselves in anomalous many-body revivals starting from a period-3 density wave initial states. 

In order to understand the properties of quantum many-body scars, in Sec.~\ref{sec:fsa} we formulate the forward scattering approximation (FSA), originally introduced in Ref.~\onlinecite{Turner2017}. After illustrating the FSA on a toy example of a free paramagnet, we describe in detail the approximate construction of scarred eigenstates, and demonstrate that the FSA method can be efficiently implemented in large systems using techniques of matrix product states. This allows one to accurately capture even non-local properties of quantum scarred eigenstates such as the entanglement entropy. 

Finally, in Sec.~\ref{sec:pert} we investigate the stability of quantum scars to various perturbations to the considered model. Using the intuition provided by the FSA, we classify perturbations according to how effective they are in destroying the quantum scarred eigenstates. We show that perturbations that are most effective in destroying the band of scarred eigenstates are also the ones that lead to the fastest thermalization according to the ETH. We conclude with the summary of main results and a discussion of open questions in Sec.~\ref{sec:disc}. Various technical details are delegated to the Appendices. 

\section{Kinetically constrained PXP model}\label{sec:model}

In this Section we start with the derivation of the effective Hamiltonian from the microscopic description of the Rydberg atom chain. Next, we consider the Hilbert space structure of the constrained model. Finally, we discuss the symmetries of the model. 
\subsection{Derivation of effective Hamiltonian \label{Sec:Ham}}
The microscopic Hamiltonian describing a chain of Rydberg atoms~\cite{Schauss2012,Labuhn2016,Bernien2017} is given by
\begin{equation}\label{eq:fullH}
  H = \sum_{j=1}^L \left(\frac{\Omega}{2} X_j - \Delta Q_i\right) + \sum_{i<j}^L V_{i,j} Q_i Q_j, 
\end{equation}
where $\Omega$ is the Rabi frequency, $\Delta$ is the detuning parameter, and $V_{i,j}\propto 1/|i-j|^6$ is the van der Waals interaction between the atoms.
We assume that each atom can be either in the ground state ($\ket{\circ}$) or in a particular excited state ($\ket{\bullet}$), thus the effective Hilbert space is that of $L$ spin-1/2 degrees of freedom.
The Rabi term is represented by the Pauli operator $X_j$ which flips the atom at site $j$ between $\ket{\circ}$ and $\ket{\bullet}$ states. The diagonal term $Q_j=(\mathds{1} + Z_j)/2$ is given by the Pauli $Z$ matrix and corresponds to the density of excitations on a given site. Note, this density is not conserved, and atoms interact with each other only when they are in the excited state. It will be convenient to introduce the projector $P_j$ onto $\ket{\circ}$ state at site $j$:
\begin{equation} \label{eq:P}
  P_j \equiv |{\circ}_j\rangle \langle {\circ}_j | = \frac{\mathds{1} - Z_j}{2}.
\end{equation}
This projector corresponds to the local density of atoms in the ground state. It is related to the density of excitations as $Q_j = \mathds{1} - P_j$ and obeys $X_j P_j = Q_j X_j$.
 
We are interested in the limit of strong nearest-neighbour interactions which we denote as $V=V_{i,i+1}$, $V\gg \Omega$, and we set $\Delta=0$, unless specified otherwise. 
Rescaling the Hamiltonian by the (inverse) nearest-neighbour interaction strength, $1/V$, and introducing the small parameter $\epsilon = \Omega / (2 V)$, we obtain the Hamiltonian
\begin{equation}\label{Eq:Hfull}
  H = H_0 + \epsilon H_1 = \sum_j Q_j Q_{j+1} + \epsilon \sum_j X_j \text{.}
\end{equation}
The dominant term $H_0$ counts the number of adjacent excitations; accordingly, its eigenvalues are the non-negative integers and they are highly degenerate.
The perturbation term $H_1$ is the trivial paramagnet; its eigenvalues are every other integer between $-L$ and $+L$ inclusive, where $L$ is the number of atoms.

In the limit of strong interactions (small $\epsilon$), we derive the effective Hamiltonian via the Schrieffer-Wolff transformation~\cite{Schrieffer:1966hu,Bravyi:2011fg}.
First, we introduce the low-energy subspace spanned by configurations with  no adjacent excited states. The projector onto this subspace can be written as 
\begin{equation} \label{Eq:proj}
  \mathcal{P} = \prod_j (\mathds{1} - Q_jQ_{j+1}).
\end{equation}
Since the leading part of the Hamiltonian~(\ref{Eq:Hfull}) vanishes in this subspace, we must consider the first non-trivial order that is given by $H_\text{SW}=  \epsilon \mathcal{P} H_1 \mathcal{P}$. Removing the overall scale $\epsilon$, we obtain the resulting effective ``PXP" model,
\begin{equation}\label{eq:H_PXP}
  H = \sum_{j=1}^L P_{j-1} X_j P_{j+1}, 
\end{equation}
with the effective constraint that no two excitations may be adjacent. This model will be the focus of Secs.~\ref{sec:eth}-\ref{sec:fsa}. Periodic boundary conditions (PBC) are imposed in the usual manner by identifying atoms $L+1$ and $1$, while for open boundary conditions (OBC) we add boundary terms $X_1 P_2$ and $X_{L-1} P_L$.

\begin{figure}[tb]
  \centering
  \includegraphics[width=\columnwidth]{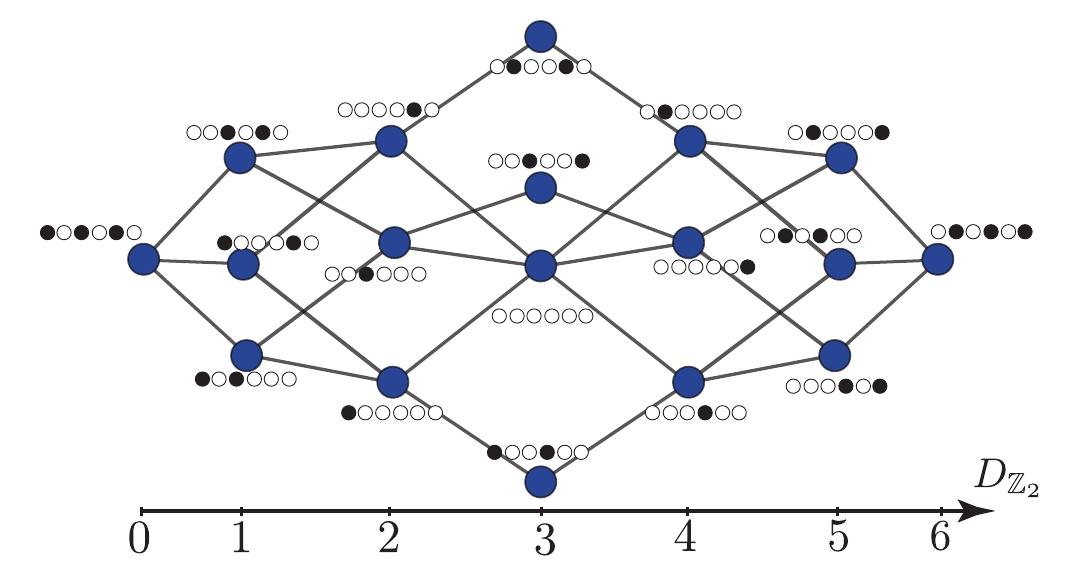}
 \caption{Graph representation of the PXP model with $L=8$ sites. The vertices of the graph are labelled by product state configurations of the atoms, where $\circ$ denotes an atom in the ground state and $\bullet$ is an atom in the excited state. Edges connect those configurations that map into each other under the action of the Hamiltonian. Horizontal axis shows the minimal number of excitations required to reach the N\'eel state from any given vertex, which coincides with the Hamming distance, $D_{\mathbb{Z}_2}$.}
  \label{fig:hamming_diagram}
\end{figure}

\subsection{Structure of the Hilbert space and symmetries \label{Sec:Hilbert}}

 The constraint on the dynamics in Eq.~(\ref{eq:H_PXP}) is rather unusual as it  destroys the tensor product structure of the Hilbert space. Restricting to the lowest-energy subspace defined by the projector $\cal P$ in Eq.~(\ref{Eq:proj}) amounts to excluding configurations with two adjacent excitations,~${\bullet}{\bullet}$. Considering the simplest example of two sites, it is clear that  the lowest-energy subspace, spanned by configurations $\{ {\circ}{\circ},{\bullet}{\circ},{\circ}{\bullet}\}$, cannot be obtained as a tensor product of local on-site Hilbert spaces. We note that, similar to other kinetically constrained models, in the lowest-energy sector the model~(\ref{Eq:proj}) has a ``flat" potential energy landscape.

It is easy to see that the Hilbert space dimension of the PXP model in Eq.~(\ref{eq:H_PXP}) grows according to the Fibonacci sequence.
Indeed, we first note that configurations of the constrained Hilbert space of an open system of $L$ sites can end with either ${\bullet}$ or ${\circ}$. A configuration ending with a ${\bullet}$ state can be obtained by appending  ${\circ}{\bullet}$ to a specific product state of a system with $L-2$ sites. On the other hand, a configuration ending with ${\circ}$ can be obtained by adding ${\circ}$ to a specific configuration of a system with $L-1$ sites. Therefore, the dimension of the Hilbert space with $L$ sites, $d_L$, satisfies the linear recurrence equation:
\begin{equation}
  d_L = d_{L-1} + d_{L-2},
\end{equation}
with initial conditions $d_0=1$, $d_1=2$. This is the well-known Fibonacci recurrence, hence $d_L$ for OBC coincides with $(L+2)$th Fibonacci number, $d_L=F_{L+2}$. The Hilbert space for $L$ sites with PBC can be formed by taking the Hilbert space of the same number of sites with open boundary conditions and removing all configurations which both begin and end with ${\bullet}$, hence $d^\text{PBC}_L = d_L - d_{L-4} = F_{L-1}+F_{L+1}$.

Finally, as illustrated in Fig.~\ref{fig:hamming_diagram}, the Hilbert space and the Hamiltonian of the PXP model have a useful graph representation. It is instructive to first consider the free paramagnet Hamiltonian, $H_\text{PM} = \sum_i X_i$,  acting in the full Hilbert space of $2^L$ product state configurations. Such a Hilbert space and the action of the Hamiltonian can be conveniently represented as a $L$-dimensional hypercube. All vertices of the hypercube can be uniquely labelled by product state configurations, e.g., $\{ {\circ}{\circ},{\bullet}{\circ},{\circ}{\bullet},{\bullet}{\bullet}\}$ for $L=2$, and edges connect configurations that differ by the state of any single atom. The Hamiltonian of the free paramagnet $H_\text{PM}$ is formally equal to the adjacency matrix of the hypercube graph.  

Next, we can consider the action of the PXP Hamiltonian in the full Hilbert space discussed above. Due to projectors dressing the $X$ operator in Eq.~(\ref{eq:H_PXP}), the state of a given atom can be flipped only if its nearest neighbours are both in the ${\circ}$ state. Thus, any local two-site configuration, $\hdots{\bullet}{\bullet}\hdots$, is frozen regardless of the state of other atoms, and a hypercube graph splits into a number of disjoint components. The largest of these components contains the state $\ket{{\circ}{\circ}{\circ}\ldots{\circ}{\circ}}$ and coincides with the subspace defined by the projector $\cal P$ in Eq.~(\ref{Eq:proj}). Notably, this implies that the Hilbert space of the PXP Hamiltonian can be viewed as a \emph{subgraph} of the $L$-dimensional hypercube. This graph is known in the mathematical and computer science literature under the names of Fibonacci~\cite{FibCube} and Lucas cube~\cite{LucCube} for OBC and PBC, respectively. 

Fig.~\ref{fig:hamming_diagram} shows the graph representation of the constrained Hilbert space and PXP Hamiltonian for $L=8$ sites with PBC. Vertices of the graph are classical product states of atoms, which have been arranged according to the Hamming distance from the particular product state, $\ket{{\bullet}{\circ}{\bullet}{\circ}{\bullet}\ldots}$, where atoms on odd numbered sites are in the excited state. This representation of the Hilbert space will play a crucial role in the forward scattering approach in Sec.~\ref{sec:fsa}.

Finally, we consider symmetries of the PXP model in Eq.~(\ref{eq:H_PXP}). Restriction to a particular symmetry sector allows to reach larger system sizes in exact diagonalization which will be used below. Moreover it is also crucial for the study of thermalization of  eigenstates. The PXP model has a discrete spatial inversion symmetry $I$ which maps site $j \mapsto L-j+1$. With PBCs, the PXP model also has translation symmetry. In, addition, the existence of the operator ${\cal C} = \prod_i Z_i$ anticommuting with the Hamiltonian~(\ref{eq:H_PXP}) leads to the particle-hole symmetry of the many-body spectrum: each eigenstate at energy $E$ has a partner at energy $-E$.

Unless specified otherwise, our results below are for PBCs where translational and inversion symmetries are explicitly taken into account. This allows us to obtain the complete set of eigenstates of large systems of up to $L=32$ sites (at this system size, the zero momentum inversion-symmetric sector of the Hilbert space contains ${\cal D}_{0+}=77436$ states).  Shift-invert algorithm allows us to extract a subset of eigenstates for larger systems of up to $L=36$ sites with ${\cal D}_{0+}=467160$.

\section{Thermalization and entanglement of eigenstates}\label{sec:eth}

In this Section, we investigate properties of eigenstates of the model in Eq.~(\ref{eq:H_PXP}) using exact diagonalization and shift-invert algorithm~\cite{SLEPc}. 
First, we directly test the ETH using the diagonal and off-diagonal matrix elements of local observables between the system's eigenstates. We find that the  majority of the eigenstates appear thermal; however, the convergence of the observables to the value dictated by the Gibbs ensemble is found to be parametrically slower compared to the ETH predictions.  In addition, we find a small number of ``special" eigenstates that strongly violate the ETH. Further, we study the eigenstate entanglement entropy, finding that the majority of states follow the usual scaling of entanglement entropy with the volume of the subsystem. In contrast, the special eigenstates exhibit a small, sub-thermal amount of entanglement. 

Finally, we show that the special eigenstates, identified as violating the ETH and having low entanglement entropy, in fact coincide with the anomalous eigenstates which are responsible~\cite{Turner2017} for the many-body revivals observed experimentally. The number of special eigenstates scales algebraically with the system size $L$, while the total number of eigenstates scales exponentially. Nevertheless, the special eigenstates are of physical significance owing to their high overlap with the simple charge-density-wave product states, which have been prepared in experiment~\cite{Bernien2017}.  While our previous paper~\cite{Turner2017} focused on the N\'eel~($|\mathbb{Z}_2\rangle$) initial state and anomalous states which have a high overlap with that state, here we establish another set of special eigenstates which are distinguished by their overlap with $\mathbb{Z}_3$ product state. These special states give rise to a different pattern of many-body revivals for the dynamics initialized in $|\mathbb{Z}_3\rangle$ state.

\begin{figure*}[t]
\begin{center}
  \includegraphics[width=0.33\textwidth]{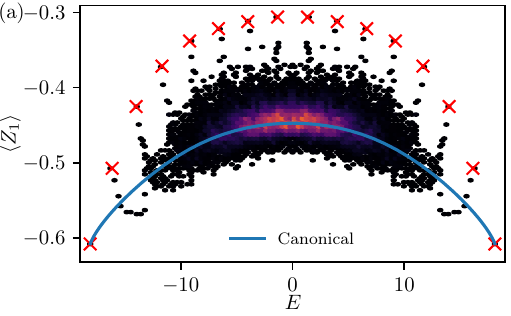}
  \includegraphics[width=0.33\textwidth]{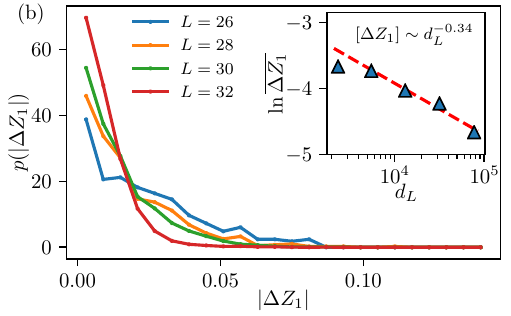}
  \includegraphics[width=0.3\textwidth]{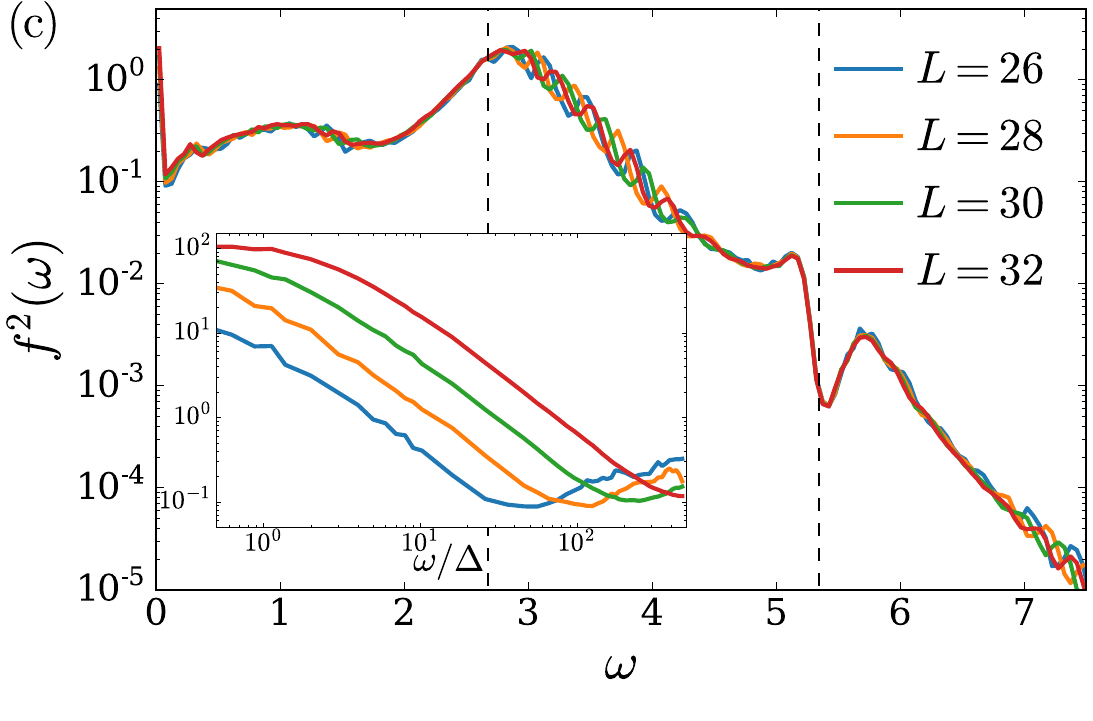}
  \caption{ \label{fig:eth} 
      (a) Strong violation of the ETH revealed by the eigenstate expectation values $\langle O^Z \rangle \equiv \langle Z_1 \rangle$, plotted as function of energy (color scale indicates the density of data points). While the majority of points are concentrated in the vicinity of the canonical ensemble prediction,  the band of special eigenstates (indicated by crosses) is also clearly visible. For these eigenstates,  $\langle O^Z \rangle $ strongly deviates from the canonical prediction at the corresponding energy. The system contains $L=30$ atoms in the zero momentum, inversion-symmetric sector.
    (b) Probability distribution for the difference in expectation value of the local observable $O^Z$ between eigenstates adjacent in energy. 
   Inset: Mean $\overline{\Delta O^Z}$ decays with a power $\approx 1/3$ of the Hilbert space dimension $d^\text{PBC}_L$ as the system size is increased (shown up to $L=32$). Averaging is performed over eigenstates in an interval between adjacent special states in the middle of the spectrum. The line shown is a linear regression to the three largest system sizes.
   (c) Off-diagonal matrix elements are a smooth function of the energy difference. Moreover, $f^2(\omega)$ does not depend on the system size, consistent with the ETH. At the same time, a number of features are visible in $f^2(\omega)$ at the frequency coinciding with the energy separation of special eigenstates in panel (a). The inset shows that $f^2(\omega)$, plotted as a function of energy in units of many-body level spacing $\Delta$, does not have a well-developed plateau until $\omega\leq \Delta$. 
  }
\end{center}
\end{figure*}

\subsection{Breakdown of ETH in special eigenstates \label{sec:eth-breakdown}}

Thermalization in ergodic systems is explained by the powerful conjecture regarding the nature of eigenstates --- the eigenstate thermalization hypothesis (ETH)~\cite{DeutschETH,SrednickiETH,RigolNature}.
The ETH states that in ergodic systems, the individual excited eigenstates have thermal expectation values of physical observables, which are identical to those obtained using the microcanonical and Gibbs ensembles.
The expectation value of a physical observable associated with an operator $O$ is given by the diagonal matrix element $O_{\alpha\alpha}= \bra{\alpha}O\ket{\alpha}$, where $\ket{\alpha}$ is an eigenstate of $H$, $H\ket{\alpha}=E_{\alpha}\ket{\alpha}$. 
Further, to describe how the system approaches the thermal state, Srednicki introduced an ansatz for the matrix elements of physical operators in the basis of system's eigenstates~\cite{Srednicki96,Srednicki99}:
\begin{equation}
  \label{eq:ETH}
  O_{\alpha\beta}={\mathcal O} (E)\delta_{\alpha\beta} +e^{-S(E)/2} f(E,\omega) R_{\alpha\beta}. 
\end{equation}
The first term describes the diagonal part of the operator in the eigenstate basis, and ${\mathcal O} (E)$ is a smooth function of energy that coincides with the canonical ensemble prediction.
The second term describes off-diagonal matrix elements, where $S(E)$ is the thermodynamic entropy at the average energy $E=(E_{\alpha}+E_\beta)/2$, and $f(E,\omega)$ is a smooth function of $E$ and the energy difference $\omega=E_{\alpha}-E_\beta$.
Finally, $R_{\alpha\beta}$ is a random number with zero mean and unit variance. We note  that the ETH ansatz~(\ref{eq:ETH}) for the matrix elements has been verified in several low-dimensional models~\cite{Rigol-FDT,Prelov13,Haque15,Alessiorev},  while it was found to break down in many-body localized systems~\cite{Serbyn15,Serbyn-17}.

In Fig.~\ref{fig:eth}(a) we test the ansatz~(\ref{eq:ETH}) for the diagonal matrix elements of the operator $O^Z = ({1}/{L}) \sum_{j=1}^L Z_j $ in the PXP model in Eq.~(\ref{eq:H_PXP}). With translation symmetry, this is equivalent to the expectation value of the $Z$ operator on the first site, $\langle Z_1 \rangle$. Moreover, due to the existence of the Hilbert space constraint, the operator $O_Z$ can be related to the nearest neighbour correlation function $O^{ZZ} = ({1}/{L}) \sum_{j=1}^L Z_j Z_{j+1}$.
Note that, because of the constraint, all eigenstates have negative values for $\langle O^Z \rangle$, rather than  $\langle O^Z \rangle \approx 0$ which would be expected in a generic thermalizing system with an unconstrained Hilbert space. Fig.~\ref{fig:eth}(a) shows that most of the expectation values of $O^Z$ are close to the canonical prediction, ${\mathcal O} (E)$, which is calculated from the Gibbs states defined by the density matrix $ \rho \propto \exp (-\beta H)$. The value of $\beta \in (-\infty,+\infty)$ is extracted by relating the observable expectation value to the mean energy in the Gibbs ensemble. However, Fig.~\ref{fig:eth} also shows that there is a number of special states that clearly violate the ETH. These states (denoted by crosses) form a distinct band, which includes the ground state of the system and extends all the way up to the middle of the spectrum. The number of states in this band is $L+1$ for OBC. For the case of systems with even $L$ and PBC, there are $L/2+1$ states in zero-momentum sector and $L/2$ states in $\pi$-momentum sector, resulting in the same total count. The special eigenstates belonging to this band can be viewed as parent states that define the ETH-breaking ``towers", visible in Fig.~\ref{fig:eth}(a). Lower states in the towers also break the ETH, though more weakly.

In Fig.~\ref{fig:eth}(b) we show the distribution of differences in the expectation value of $O^Z$ between eigenstates adjacent in energy, $\Delta O_i^Z = |O^Z_{i+1,i+1}-O^Z_{ii}|$. Consistent with the ETH prediction, we observe that this distribution narrows around $\Delta O^Z = 0$ upon increasing the system size.  However, despite fluctuations of $\Delta O^Z$ decaying with the system size, this decay is parametrically slower compared to the standard ETH prediction. The inset of Fig.~\ref{fig:eth}(b)  shows that the mean $\overline{\Delta O^Z}$ decays approximately as $1/{\cal D}_{0+}^{1/3}$ whereas the ETH ansatz (\ref{eq:ETH}) would suggest a decay which is inversely proportional to the square root of the density of states, $1/\sqrt{\cal D}_{0+}$.  A recent study~\cite{Khemani2018} of the same model with OBC also reports the scaling of diagonal matrix elements to be slower than expected from the ETH.
Note, however, that only the few largest system sizes in Fig.~\ref{fig:eth}(b) appear to be in the scaling regime, which means that it is possible that the power governing the decay of the diagonal matrix element converges to $1/2$ in  larger systems. 

Finally, we test the ETH ansatz for the off-diagonal matrix elements. Using Eq.~(\ref{eq:ETH}) we define the average matrix element at a given energy separation,
\begin{equation}\label{Eq:f2}
f^2(\omega)  = e^{S(E)} \langle |\bra{\beta}O^Z\ket{\alpha}|^2 \delta(E_\alpha-E_\beta-\omega)\rangle_{\alpha,\beta},
\end{equation}
which is rescaled by the density of states. In what follows, we refer to $f^2(\omega)$ as the infinite temperature spectral function, since averaging in Eq.~(\ref{Eq:f2}) is performed over the middle 2/3 eigenstates in the spectrum, denoted by ${\alpha,\beta}$. If the off-diagonal matrix elements obey the ETH, the function $f^2(\omega)$ ought to be smooth and independent of the system size. This is indeed confirmed by Fig.~\ref{fig:eth}(c), which shows the collapse of $f^2(\omega)$ for different system sizes. With the previously chosen normalization for the operator $O^Z$, in Fig.~\ref{fig:eth}(c) we have multiplied $f^2(\omega)$ by $L$, which yields the best collapse of the curves within the available system sizes.~\cite{Alessiorev,Rigol17}  Moreover, $f^2(\omega)$ decays exponentially at large $\omega$, as expected from the locality of the Hamiltonian~\cite{SlowHeating,Alessiorev}.

Surprisingly, in the intermediate range of frequencies we observe non-monotonic behavior of $f^2(\omega)$. The positions of the characteristic features in $f^2(\omega)$ coincide with the energy separation between the ETH-breaking eigenstates  in Fig.~\ref{fig:eth}(a). Such a behavior, to the best of our knowledge, has not been reported before in the context of translationally invariant systems without disorder.~\cite{Rigol17} (Note that Ref.~\onlinecite{Alessiorev} observed features in the spectral function at energies $O(1/L)$ for a system of hard-core bosons with dipolar interactions in a harmonic trap that breaks translational invariance.) In contrast, in disordered systems, the emergence of a similar peak was interpreted as a signature of local resonances~\cite{Serbyn-17}. In addition, the inset of Fig.~\ref{fig:eth}(c) shows that $f^2(\omega)$ does not have a well-developed plateau until $\omega$ becomes of the order of the many-body level spacing, $\Delta\propto \sqrt{L}/{\cal D}_{0+}$. Such a plateau is typical for thermalizing systems, and it sets the energy scale (the Thouless energy) below which the system essentially can be described by a random matrix ensemble~\cite{Alessiorev}. 

From the absence of saturation in the matrix elements at small energies, we expect the level statistics to show deviations from the Wigner-Dyson form. Indeed, previously it was demonstrated~\cite{Turner2017} that for small system sizes $L\leq 28$ the level statistics is approximately described by the Semi-Poisson distribution~\cite{Bogomolny99}. This is consistent with the approximately critical form of $f^2(\omega)$ for $\omega\geq \Delta$ in Fig.~\ref{fig:eth}(c)~\cite{Serbyn16,Serbyn-17}. In addition, we also expect the level compressibility to be enhanced compared to the Wigner-Dyson ensemble. However, the slow development of the plateau for $L\geq 30$ suggests that both the level statistics and compressibility approach the Wigner-Dyson ensemble for larger system sizes. 

The absence of a Thouless plateau in the off-diagonal matrix elements, along with the slow decay of fluctuations in diagonal matrix elements, $\overline{\Delta O^Z}$, and deviations from purely Wigner-Dyson level statistics,  suggests that thermalization of the bulk of eigenstates in the PXP model may not fully follow the ETH. We return to the discussion of thermalization in Sec.~\ref{sec:pert}. There we will show that full thermalization is restored, and the system follows the canonical ETH predictions, once the PXP model is perturbed in  a way that fully destroys the special bands of eigenstates. 

\subsection{Entanglement of eigenstates}

Quantum entanglement is a complementary probe of thermalization and its breakdown, which provides additional insights compared to matrix elements of physical observables. Equivalence of all observables to their canonical values imposed by the ETH implies that the von Neumann entanglement entropy of a subregion $A$ in an eigenstate $\alpha$, $S^\alpha = -{\rm tr}_A \left( \rho_A^\alpha \ln \rho_A^\alpha \right)$, is equal to the thermodynamic entropy of $A$ at temperature $T$ which corresponds to the eigenstate energy $E_\alpha$. Here, the entanglement of an eigenstate is defined in terms of its reduced density matrix $\rho_{A}^\alpha={\rm tr}_B \ket{\alpha} \bra{\alpha}$ that is obtained by tracing out the degrees of freedom in the complement of the spatial region $A$, denoted as $B$. Thermodynamic entropy scales proportionally to the volume of region $A$ and is maximal in the middle of the band where the density of states is highest. 

\begin{figure}[tb]
  \begin{center}
  \includegraphics[width=0.95\columnwidth]{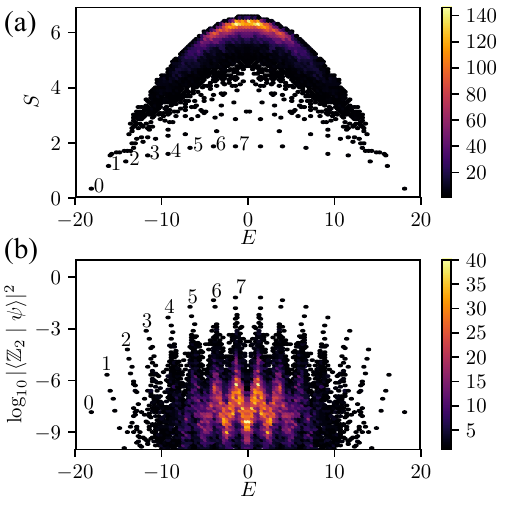}
  \caption{\label{fig:entropyscatter}
    (a) Bipartite entanglement entropy of eigenstates, $S$, as a function of energy $E$. Region $A$ is chosen as one half of the chain. The bulk of the states have large volume-law entropy ($S\gtrsim 5$), however some outliers with anomalously low entropy ($S\lesssim 2$) are also visible. These states are labelled by $0,\ldots, 7$, and they span the entire energy range between the ground state (state $0$) and the middle of the band (state $7$).     
    (b) Density plot showing the joint distribution of energy and overlap with $\ket{\mathbb{Z}_2}$ product state among the energy eigenstates.
    The states with largest overlap are identified with the low entropy states from the top panel.
    Data shown is for $L=30$ sites in the zero-momentum and inversion symmetric sector.
  }
  \end{center}
\end{figure}

Fig.~\ref{fig:entropyscatter}(a) shows that entanglement entropy $S$ for the majority of eigenstates exhibits behavior that is consistent with the predictions of the ETH. Finite-size scaling of states with large entropy ($S\gtrsim 5$) reveals volume-law scaling, $S\propto L$ (not shown). However, in addition to the bulk of typical highly-entangled states, we also observe outliers with much lower entropy. The outlier states with the lowest entanglement, labeled as $0,\ldots,7$ in Fig.~\ref{fig:entropyscatter}(a), span the entire bandwidth. Note that we do not label states at $E>0$, as they are related to states $0,\ldots,7$ by particle-hole symmetry. 

For even system size $L$, there are $L/2 +1$ special eigenstates in the zero momentum sector, and $L/2 -1$ such states in $\pi$-momentum sector. Thus in total, we observe $L+1$ special states. These states coincide with the states that maximally violate the ETH, depicted by crosses in Fig.~\ref{fig:eth}(a). Furthermore, as shown in Ref.~\onlinecite{Turner2017}, and as we discuss in more detail in the following Section, these special states can also be identified as ones that have highest overlap with $|\mathbb{Z}_2\rangle$ product state defined in Eq.~(\ref{eq:cdw}) below, as illustrated in Fig.~\ref{fig:entropyscatter}(b).

In Sec.~\ref{sec:fsa} we present an approach based on forward scattering, which accurately captures the  highly excited eigenstates with low-entropy labeled in Fig.~\ref{fig:entropyscatter}. (A brief account of this approach was presented in Ref.~\onlinecite{Turner2017}.) Within the forward scattering approximation, we will be able to demonstrate that these special eigenstates are highly atypical from the entanglement point of view: their entanglement entropy scales with the \emph{logarithm} of system size, i.e., $S\propto \ln L$. This type of behavior, which is very different from the ETH prediction, is commonly encountered in ground states of critical systems~\cite{Calabrese2004} and systems with Fermi surfaces~\cite{KlichFL, CalabreseFL}. Similar phenomenology is found in recent work~\cite{Bernevig2017, BernevigEnt}, where exact expressions for special excited eigenstates in the non-integrable AKLT model were found.

\subsection{Overlap of special eigenstates with product states\label{sec:product}}

We have demonstrated that the PXP model breaks the ETH because of the existence of a relatively small (algebraic in the system size) number of highly atypical, non-thermal eigenstates. These states are distinguished by anomalous matrix elements of local observables, Fig.~\ref{fig:eth}(a), as well as by sub-thermal entanglement entropy, Fig.~\ref{fig:entropyscatter}(a). However, there exist only $L+1$ such states embedded among an exponentially many (slowly) thermalizing eigenstates. Hence, naively one may expect that these states do not have direct physical relevance, as they might be hidden by the contribution of a much larger number of typical eigenstates.  Below we show that this is not the case because special eigenstates have anomalously high overlaps with certain product states. This implies that superpositions of special eigenstates can be experimentally prepared and probed using a global quench. For example, a class of product states which was studied in recent experiments~\cite{Bernien2017} are the charge density wave (CDW) states 
\begin{equation}\label{eq:cdw}
  |\mathbb{Z}_k\rangle = |\ldots \underbrace{{\bullet} {\circ}\ldots{\circ}}_k{\bullet}\ldots\rangle,
\end{equation}
where the atoms in the excited state are separated by $k-1$ atoms in the ground state. In this Section we show that  the simplest CDW states, the period-2 ($\mathbb{Z}_2$ or N\'eel) state and the period-3 ($\mathbb{Z}_3$) state, allow one to identify a dominant subset of special states in the PXP model.

Fig.~\ref{fig:entropyscatter}(b) shows the squared overlap between all the eigenstates of the PXP model and $\ket{\mathbb{Z}_2}$ product state on the logarithmic scale. From this plot, we see that there exists a set of eigenstates with anomalously large overlap, which form regular tower structures. The states at the top of towers coincide with the special eigenstates identified via the breakdown of the ETH in Fig.~\ref{fig:eth}(a) and entanglement entropy in Fig.~\ref{fig:entropyscatter}(a). We also see that for each of the special states labeled $0,\ldots 7$, there are further eigenstates belonging to the same tower (i.e., with similar eigenenergy), which have much larger overlap with the N\'eel state compared to the majority of thermalizing states. 

\begin{figure}[b]
  \begin{center}
  \includegraphics[width=0.95\columnwidth]{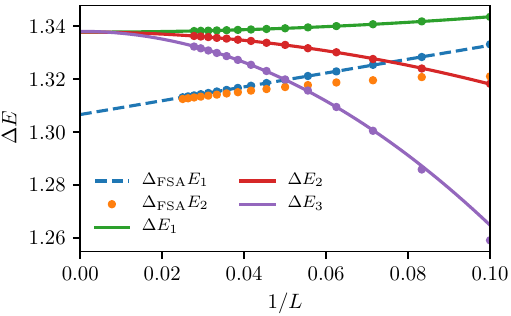}
  \caption{ \label{Fig:gaps}
    Finite size scaling of the energy gaps between special states closest to the middle of the spectrum shows the convergence of all gaps to the same value in the thermodynamic limit. Note that adjacent special states belong to different momentum sectors.  The gaps accurately follow quadratic dependence  on  $1/L$ in the range of available system sizes $L=10,\ldots, 36$. The prediction of the forward scattering approximation, discussed in Sec.~\ref{sec:fsa}  for systems with up to $L=48$ sites, is shown by blue/orange points, corresponding to the two states closest to the middle of the spectrum. Within this approximation, the energies appear to follow linear dependence in $1/L$ (dashed line). 
  }
  \end{center}
\end{figure}

 Interestingly, the $L/2+1$ special eigenstates from the zero-momentum sector, half of which are highlighted in Fig.~\ref{fig:entropyscatter}, are nearly equidistant in energy. Near the center of the many-body band, they are separated in energy by $\Delta E \approx 2.66$.  In addition, the $L/2-1$ special states from the $\pi$-momentum inversion-antisymmetric sector have energies exactly between the special states from the zero-momentum sector. Thus, combining both sectors, the energy separation between special states becomes $\Delta E \approx 1.33$ in the middle of the spectrum. Fig.~\ref{Fig:gaps} shows the finite-size scaling of the energy gaps between the four special eigenstates closest to the energy $E=0$. All the gaps accurately follow the finite size scaling $\Delta E_i = 1.337 + c_i/L^2$, where a linear term is absent. Constants $c_i$ depend on the chosen pair of eigenstates, with $c_1=0.582$ corresponding to the gap between the special state at $E=0$ and the closest one with non-zero energy.  In contrast, the distance between special eigenstates at the edge of the spectrum, e.g., the ground state (0th special eigenstate), which always belongs to the zero-momentum sector, and the first special eigenstate that lives in $\pi$-momentum sector is $\Delta E_0 \approx 0.96$. This behavior should be contrasted with the AKLT model,~\cite{Bernevig2017} where the special excited states are equidistant in energy.

\begin{figure}[tb]
  \includegraphics[width=0.999\columnwidth]{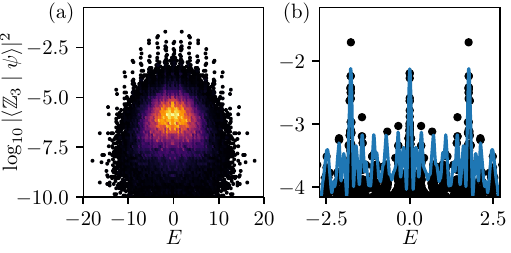}
  \caption{ \label{fig:z2z3overlap}
    (a) The overlap between eigenstates of the PXP model and $|\mathbb{Z}_3\rangle$ state as a function of energy $E$ reveals another special band of eigenstates.
    (b) Same plot but zoomed in around energy $E=0$. Black dots mark individual eigenvalues, while the blue curve is a Gaussian convolution of the overlap probability $|\langle\mathbb{Z}_3|\psi(E)\rangle|^2$ viewed as a function of energy. This reveals a number of peaks subdividing the interval between the highest overlap states.
    Both plots are for $L=30$ in the zero-momentum and inversion symmetric sector together with the $\pm{}2\pi/3$ momentum sectors.
  }
\end{figure}

Finally, we note that in addition to the special states identified via the overlap with $|\mathbb{Z}_2\rangle$ state, there are further states that also violate the ETH but more weakly. To identify some of them, in Fig.~\ref{fig:z2z3overlap}(a) we plot the overlap of PXP eigenstates with $|\mathbb{Z}_3\rangle$ product state. Here we can also observe the existence of a band of states with anomalously high overlap. In contrast  to the $\mathbb{Z}_2$ case, this band is less clearly separated from the bulk of the spectrum. A natural question is whether the set of special states revealed by $|\mathbb{Z}_2\rangle$ intersects with that of $|\mathbb{Z}_3\rangle$. Cross comparison of overlaps (not shown) reveals that these two sets of special states are different from each other. Zooming in on the overlap plot around energy $E=0$, shown in Fig.~\ref{fig:z2z3overlap}(b), we can observe several ``mini-towers" between the highest overlap states. This feature will give rise to more complicated dynamics in the $\mathbb{Z}_3$ case, which is discussed in the following Section.

\subsection{Dynamical signatures of special eigenstates \label{sec:dynamics}}

Anomalously high overlaps of special eigenstates with product states like $|\mathbb{Z}_2\rangle$ or $|\mathbb{Z}_3\rangle$ make them amenable to a simple experimental probe -- global quench.   In particular, the quench from $|\mathbb{Z}_2\rangle$ state was  studied experimentally in Ref.~\onlinecite{Bernien2017} and in numerical simulations on small systems~\cite{Sun2008,LesanovskyDynamics,Olmos2012}.  We initialize the system at time $t=0$ in the state $|\psi(0)\rangle = |\mathbb{Z}_k\rangle$, and then follow the evolution of this initial state with the PXP Hamiltonian, Eq.~(\ref{eq:H_PXP}), $\ket{\psi(t)} = \exp(-iHt)\ket{\psi(0)}$.
This evolution is determined by how $|\psi(0)\rangle$ is decomposed in terms of the system's eigenstates.
 
Figs.~\ref{fig:entropyscatter}(b) and \ref{fig:z2z3overlap}(a)  demonstrate that there are a few eigenstates with high overlaps \emph{and} constant energy separation in the middle of the band where the overlaps are largest (see Fig.~\ref{Fig:gaps}). Therefore, we expect that quantum quench from $|\mathbb{Z}_2\rangle$ or $|\mathbb{Z}_3\rangle$ product state will give rise to coherent oscillations, with a frequency determined by the energy separation between the towers of special states in Fig.~\ref{fig:entropyscatter} or Fig.~\ref{fig:z2z3overlap}. These oscillations in the dynamics can be observed by measuring the expectation values of certain local observables~\cite{Bernien2017,Turner2017}, or more generally, using the quantum fidelity (or return probability)   defined as   $|\bra{\mathbb{Z}_k} \exp(-i H t) \ket{\mathbb{Z}_k} |^2$.

\begin{figure}[tb]
  \begin{center}
  \includegraphics[width=0.95\columnwidth]{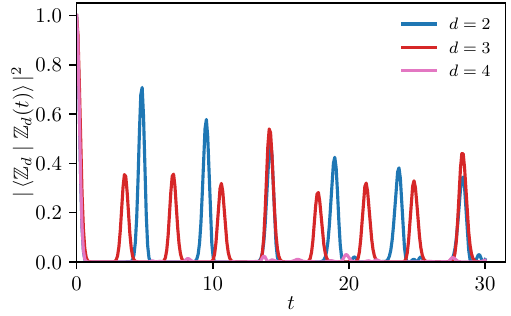}
  \caption{ \label{fig:fidelity}
    Quantum fidelity shows periodic in time revivals for $|\mathbb{Z}_2\rangle$ and $|\mathbb{Z}_3\rangle$ initial product states. In contrast,  $|\mathbb{Z}_4\rangle$ initial state shows a complete absence of revivals. Data is for system with $L=24$ sites with periodic boundary conditions.
  }
  \end{center}
\end{figure}

Fully consistent with the expectations described above,  fidelity for quenches from  $|\mathbb{Z}_2\rangle$, $|\mathbb{Z}_3\rangle$ initial states shown in Fig.~\ref{fig:fidelity} reveals pronounced periodic revivals. The period of these revivals   is given by $T_{\mathbb{Z}_2}=2\pi/\Delta E_\infty$, where $\Delta E_\infty \approx 1.33$ is the energy separation between the $|\mathbb{Z}_2\rangle$ special states. We note that revivals of a local observable -- the density of domain walls -- were found in Ref.~\onlinecite{Bernien2017} for the $|\mathbb{Z}_2\rangle$ case. The frequency of these revivals is identical to the frequency found here using quantum fidelity. By contrast, for $|\mathbb{Z}_4\rangle$ initial state, we do not observe any revivals in the fidelity. This is in agreement with the absence of anomalously high overlaps between eigenstates and $|\mathbb{Z}_4\rangle$ product state. 

The return probability in Fig.~\ref{fig:fidelity} shows that $\ket{\mathbb{Z}_3}$ initial state also exhibits many-body revivals. The period of these revivals is approximately given by $T_{\mathbb{Z}_3}\approx ({3}/{4}) T_{\mathbb{Z}_2}$. In addition to the revivals, the dynamics displays a beating pattern modulating the amplitude of the revivals. These modulations can be attributed to additional towers of special states, which are illustrated by the blue line in Fig.~\ref{fig:z2z3overlap}(b). These additional towers are situated between the highest overlap states. This secondary band of special eigenstates generally has enhanced overlaps with product states containing a domain wall between two different  $\mathbb{Z}_3$ patterns, each spanning one half of the system, e.g., ${\circ}{\circ}{\bullet}\ldots{\circ}{\circ}{\bullet}|{\circ}{\bullet}{\circ}\ldots{\circ}{\bullet}{\circ}$. The existence of such a state in a finite system requires $L$ to be divisible by 6, $L=6\ell$. These ``secondary'' special states introduce an additional frequency that is $\ell=L/6$ times smaller compared to the energy difference  between the adjacent $\ket{\mathbb{Z}_3}$ special states from all momentum sectors. Consequently, the beating pattern also appears for system sizes divisible by 6, and in Fig.~\ref{fig:fidelity} for $L=24$ we observe an enhancement of every $\ell=4$ revival.

Finally, we mention that the PXP model, in addition to special eigenstates, also exhibits an exponentially large number of states with energy $E=0$. These states can be understood as arising from the intricate interplay between the bipartite structure of the graph which describes the Hilbert space and Hamiltonian (see Fig.~\ref{fig:hamming_diagram}) and inversion symmetry present in the problem. In Appendix~\ref{sec:zero} we discuss these zero-energy states in greater detail and obtain the lower bound on their number that was reported in Ref.~\onlinecite{Turner2017}.    

In the following Section, we introduce a forward scattering method that allows us to construct accurate approximations of special eigenstates in the PXP model. Moreover, this method allows to build special eigenstates starting from $|\mathbb{Z}_2\rangle$ and $|\mathbb{Z}_3\rangle$ product states, explaining the anomalously enhanced overlaps. In addition, forward scattering will provide an insight into the different dynamical behavior of the PXP model depending on the initial state. 

\section{Forward-scattering approximation \label{sec:fsa}}

So far we have studied spectral properties of the PXP model using exact diagonalization and identified a set of special eigenstates. 
Here we explicitly construct a subset of those eigenstates which are related to $|\mathbb{Z}_2\rangle$ product state, and whose number scales linearly with the system size.  
The basic idea behind the construction of these special states is a modification of the Lanczos iteration~\cite{lanczosbook}. Below we start with applying this modification, dubbed ``forward scattering approximation'' (FSA), to the solvable example of a free paramagnet.  In this toy example, the Hilbert space and the Hamiltonian can be represented as a hypercube and its adjacency matrix, respectively (see Sec.~\ref{Sec:Hilbert}). The advantage of this toy model is that the FSA is exact. After explaining the basics of the method on this simple model, we consider the more interesting case of the PXP Hamiltonian~(\ref{eq:H_PXP}). This model differs from a free paramagnet by the projection imposed on the Hilbert space, which makes the FSA scheme approximate. We formulate the FSA scheme for the PXP model and benchmark it on a number of different properties (more detailed analysis of errors introduced by the FSA scheme can be found in Appendix~\ref{App:error}). We demonstrate that the FSA can be efficiently implemented in large systems using matrix product state methods.  Finally, in the last part of this Section we discuss the notion of a trajectory which allows us to relate special eigenstates to quantum scars in the many-body case. In addition, we discuss the implications of the FSA for the stability of special eigenstates to various perturbations of the Hamiltonian.
 
\subsection{Forward scattering on the hypercube}

We start with the FSA on the $L$-dimensional hypercube graph corresponding to the free paramagnet Hamiltonian, $H_\text{PM}=\sum_{i=1}^L X_i$. Hence there is no constraint imposed on the Hilbert space throughout this subsection. In this case the FSA method is exact, and it results in a Hamiltonian whose non-zero matrix elements are those of the spin operator $2 S^x$ for a spin of size $L/2$.  Although this result can be obtained via other means, the approach outlined here allows us to  introduce the basic ingredients that will be needed for the non-trivial case of the PXP model.  

The FSA method is a version of the Lanczos recurrence.~\cite{lanczosbook}  Lanczos recurrence is used to construct the Krylov subspace and obtain an approximation to the given Hamiltonian by its projection onto this subspace. The usual Lanczos iteration starts with a given vector in the Hilbert space, $v_0$, usually chosen to be random. The orthonormal basis is constructed by recursive application of the Hermitian matrix $H$ (i.e., the Hamiltonian) to the starting vector.  The basis vector $v_{j+1}$ is obtained from $v_j$ by applying  $H$ and orthogonalizing against $v_{j-1}$:
\begin{equation}
 \label{eqn:lanczos}
 \beta_{j+1} v_{j+1} = H v_{j} - \alpha_{j} v_{j} - \beta_{j} v_{j-1},
\end{equation}
where $\alpha_j = \left<Hv_j\mid{}v_j\right>$ and $\beta > 0$ are chosen such that $\|v_j\| = 1$. Here we observe that the action of $H$ results in the next vector $v_{j+1}$ (``forward propagation''), but also gives some weight on the previous basis vector, $v_{j-1}$ (``backward propagation'').  

In the case of the free paramagnet Hamiltonian, the above scheme can be simplified. Let us choose the specific initial vector as the N\'eel basis state $v_0 = \ket{\mathbb{Z}_2}=\ket{{\bullet}{\circ}{\bullet}{\circ}\ldots}$. Moreover, we split the Hamiltonian $H_\text{PM} = \sum_i X_i = H_++H_-$ into the forward and backward scattering operators,
\begin{subequations}\label{eqn:split}
\begin{align}\label{eqn:split1}
  H_{+} &= \sum_{j\in \text{ odd}} \sigma_j^- + \sum_{j\in \text{ even}} \sigma_j^+,
  \\ \label{eqn:split2}
  H_{-} &= \sum_{j\in \text{ odd}} \sigma_j^+ + \sum_{j\in \text{ even}} \sigma_j^-.
\end{align}
\end{subequations}
For the free paramagnet considered in this section, it can be seen that $H_+$ and $H_-$ obey the standard algebra of spin raising and lowering operators. This can be used to immediately  write down the Hamiltonian matrix. Nevertheless, we show how the same result can be obtained via a more general procedure, which can be directly generalized to the PXP model. 

Let us consider the first step of the recurrence~(\ref{eqn:lanczos}) in this case. Operator $H_-$ annihilates the state $\ket{{\bullet}{\circ}{\bullet}{\circ}\ldots}$, and we obtain the vector $\beta_1 v_1 = H_+ \ket{\mathbb{Z}_2}$, which is an equal superposition of all states with a single spin flip on top of $\ket{\mathbb{Z}_2}$, 
\begin{equation}\label{Eq:one-defect}
\beta_1  v_1 =\ket{{\circ}{\circ}{\bullet}{\circ}{\bullet}{\circ}\ldots}+\ket{{\bullet}{\bullet}{\bullet}{\circ}{\bullet}{\circ}\ldots}+\ket{{\bullet}{\circ}{\circ}{\circ}{\bullet}{\circ}\ldots}+\ldots.
\end{equation}
Hence we see that $H_+$ ensures forward propagation in this case, and the action of $H_-$ has vanished. The vector $v_1$ is automatically orthogonal to $v_0$, thus we set $\alpha_0=0$, and $\beta_1=\sqrt{L}$ by normalization. 

In the second step of the recurrence, we can observe that the action of $H_+$ on $v_1$ will produce a state containing a pair of defects atop the N\'eel state, which is thus orthogonal to both $v_{1}$, and $v_0$. On the other hand, the action of the backward-scattering part gives us the original state $v_0$, $H_- v_1 =  \beta_1 v_0$, where we explicitly used the value of $\beta_1$. In the case of a free paramagnet, one can show that
\begin{equation}\label{Eq:backward-scatter}
H_-  v_{j} =  \beta_{j} v_{j-1}
\end{equation}
holds more generally at \emph{every} step of the iteration. This allows to cancel $H_- v_{j} $ with the last term in Eq.~(\ref{eqn:lanczos}), yielding the FSA recurrence:
\begin{equation}\label{Eq:lanczosFSA}
  \beta_{j+1} v_{j+1} = H_+ v_{j},
\end{equation}
where we also omitted the $\alpha_{j} v_{j}$ term since all $\alpha_j=0$. This follows from the fact that $H_\pm$ operators change the Hamming distance from $\ket{\mathbb{Z}_2}$ state by $\pm 1$. Hence, the new state $v_{j+1}$ is always orthogonal to $v_{j}$. Moreover, by the same argument, the FSA recurrence closes after $L+1$ steps as it reaches the vector $v_{L} = \ket{\mathbb{Z}'_2}=\ket{{\circ}{\bullet}{\circ}{\bullet}\ldots}$, which is the translated N\'eel state that  vanishes under the action of $H_+$. 

Finally, using induction one can demonstrate that 
\begin{eqnarray}\label{eq:betajmain}
  \beta_j = \sqrt{j(L - j + 1)},
\end{eqnarray}
which, as anticipated, is the well-known matrix element of a spin ladder operator.
This reesults in the effective tri-diagonal matrix form of $H_\text{FSA}$ in the basis of $v_j$:
\begin{equation}
  \label{eq:tmatrix}
  H_\text{FSA} = \left(\begin{array}{ccccc}
    0 & \beta_1  & & \\
    \beta_1  & 0 & \beta_2  & \\
             & \beta_2  & 0 & \ddots \\
             &          & \ddots   & \ddots  & \beta_{L} \\
             &          &          & \beta_{L} & 0
   \end{array}\right)\text{.}
\end{equation}
Taking into account the expression for $\beta_j$, we see that this matrix coincides with the $2 S^x$ operator for a spin of size $L/2$, resulting in a set of $L+1$ equidistant energy levels. Likewise, the wave functions in the basis of $v_j$ can be obtained from the Wigner rotation matrix. 

\subsection{Forward scattering for PXP model}

Above we demonstrated how the FSA allows to find a subset of eigenstates in the case of a free paramagnet. Now we return to the problem of the constrained PXP model that is defined on the subgraph of the $L$-dimensional hypercube, where the FSA method is no longer exact. To see this, we again start the FSA from  $ v_0=\ket{\mathbb{Z}_2}$ state, and split the Hamiltonian Eq.~(\ref{eq:H_PXP}) into the forward and backward propagating parts, $H = H_{+} + H_{-}$ with 
\begin{align}\label{Eq:Hpm}
  H_{\pm} &= \smashoperator[lr]{\sum_{j \in \text{ even}}} P_{j-1}\sigma^\pm_jP_{j+1}  + \smashoperator[r]{\sum_{j\in \text{ odd}}} P_{j-1}\sigma^\mp_jP_{j+1} . 
\end{align}
Similar to the case of free paramagnet, in such a decomposition $H_+$ always increases the Hamming distance from the N\'eel state and $H_-$ always decreases it. In the Hilbert space graph in Fig.~\ref{fig:hamming_diagram}, $H_+$ always corresponds to moving from left to right. Hence, the FSA recurrence closes after $L+1$ steps once forward propagation reaches the opposite edge of the graph, $ \ket{\mathbb{Z}'_2}$.

\begin{figure}[tb]
  \centering
    \includegraphics[width=0.999\columnwidth]{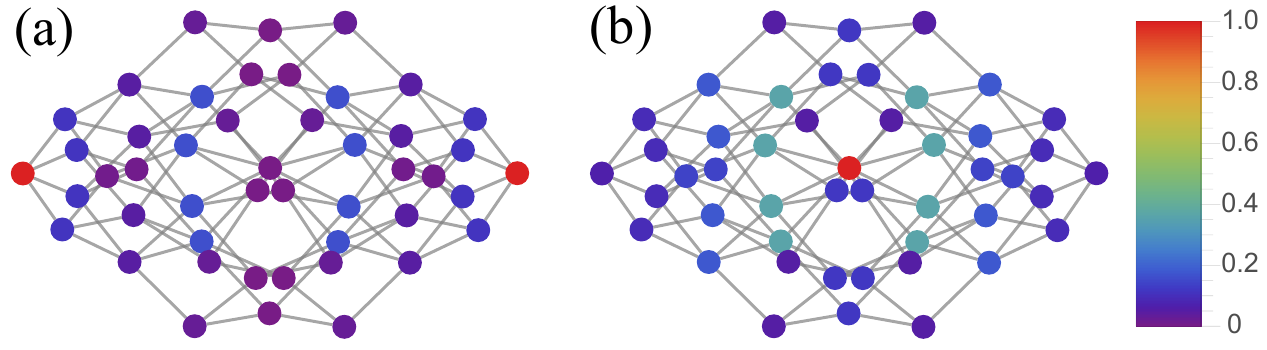}
  \caption{  \label{fig:ovl_fsa}
  Two eigenstates of the PXP model represented on the Hilbert space graph for $L=10$ sites and PBC. The color of the vertices reflects the weight of the corresponding product state in the eigenvector, where the largest weight is normalized to one. Similar to Fig.~\ref{fig:hamming_diagram}, the N\'eel states are the left/right most vertices of the graph, while the fully polarized state is located in the center of the graph.
    (a) Wave function of the special eigenstate closest to zero energy is concentrated in the vicinity of the N\'eel states. In contrast, the wave function of the ground state (b) is concentrated in the center of the graph. 
 }

\end{figure}

Now, we observe that the key property that enabled the FSA recurrence, Eq.~(\ref{Eq:backward-scatter}),  holds only approximately. More specifically, if one starts from the N\'eel state, Eq.~(\ref{Eq:backward-scatter}) is exact for $j=1,2$, but at the third step of the recurrence this property does not hold any more. Nevertheless, we can still apply the FSA recurrence as defined in Eq.~(\ref{Eq:lanczosFSA}). The error is quantified by the vector
\begin{equation}\label{eq:error}
  \delta w_j = H_{-} v_{j-1} - \beta_{j-1} v_{j-2}.
\end{equation}
The error per individual step of the FSA iteration can be shown to depend on the commutator $\left[H_+, H_- \right]$, and will be discussed in more detail in Appendix~\ref{App:error}. Generally, this error is smaller for states that are closest to the middle of the spectrum. This is because, as shown in Fig.~\ref{fig:ovl_fsa}(a), the special eigenstates closest to $E=0$ have their wave function concentrated near  the edges of the graph. As the first few steps of the FSA approximation near the edges of the graph are exact, we expect it to better capture those states that are close to zero energy. In contrast, the ground  state and other low-lying special eigenstates live primarily in the center of the graph, i.e., in the vicinity of the fully polarized state, $\ket{{\circ}{\circ}{\circ}\ldots}$, as seen in Fig.~\ref{fig:ovl_fsa}(b). 
 
The resulting vectors $v_j$ obtained from the FSA recurrence, Eq.~(\ref{Eq:lanczosFSA}), starting from $v_0 = \ket{\mathbb{Z}_2}$, form an orthonormal subspace because each is in a different Hamming distance sector and the recurrence closes after $L+1$ steps.
At present we do not have closed analytical expressions for $\beta_j$ coefficients, however they can be obtained by  a number of efficient means for systems on the order of $L\lesssim 100$ sites, as we discuss in Appendix~\ref{sec:loops}.

Diagonalizing the tridiagonal matrix of size $(L+1)\times (L+1)$ with $\beta_j$ determined either directly from Eq.~(\ref{Eq:lanczosFSA}) or via linear recurrence method explained in Appendix~\ref{sec:loops}, one can obtain a set of approximate eigenenergies and eigenvectors in the FSA basis. However, rotating the eigenvectors to the physical basis requires one to store at least $L+1$ FSA basis vectors, each of the dimension of the full Hilbert space, i.e., exponentially large in $L$. Earlier we demonstrated in Fig.~\ref{fig:entropyscatter}(a) that special eigenstates have considerably lower entropy than other eigenstates at the same energy density. This suggests that matrix product state (MPS)~\cite{Schollwock} representation of the FSA basis and special eigenstates should be highly efficient in the present case. 

In order to formulate the FSA recurrence in the MPS basis, we use the matrix product operator representation of $H_+$ from Eq.~(\ref{Eq:Hpm}) and construct the basis by applying the MPS operator to the N\'eel state. The only difference with respect to the exact FSA is that a compression similar to DMRG algorithms~\cite{Schollwock} is performed every time an operator is applied to a state or two states are summed. That is, we truncate the state for all bipartitions, so that for each reduced density matrix the truncated probability is $<10^{-8}$, and then renormalize the state. Below we discuss the physical properties of special eigenstates obtained within the FSA.

\subsection{Extracting physical properties of special eigenstates within FSA}

Diagonalizing the tridiagonal matrix with $\beta_j$ determined either directly from Eq.~(\ref{Eq:lanczosFSA}) or via linear recurrence method explained in Appendix~\ref{sec:loops}, we obtain a set of approximate eigenvectors and their energies.  Previously,  in Ref.~\onlinecite{Turner2017}  we demonstrated that eigenenergies agree within a few percent with exact diagonalization data for the largest available system of $L=32$  atoms. Here we perform a more detailed study of scaling of the FSA results. The finite size scaling in Fig.~\ref{Fig:gaps} reveals that the energy spacing between special eigenstates within the FSA approximation saturates to a value that  differs by $\approx 2.6\%$ from the one extracted from exact diagonalization. Moreover,  finite-size corrections to the FSA energy are linear in $1/L$, while exact results appear to follow $1/L^2$ corrections. The origin of this discrepancy remains to be understood. 

Moreover, earlier we reported a good agreement between the FSA eigenvectors and the projection of exact eigenvectors onto the FSA basis.~\cite{Turner2017} The FSA also correctly reproduces the expectation values of local observables. In particular, crosses in Fig.~\ref{fig:eth}(a) represent the expectation values of local observables within the FSA for a chain with $L=30$ sites. They agree very well with the exact diagonalization data. Given the ability of the FSA to capture the values of local observables, it is natural to ask if it also describes non-local properties of special eigenstates, such as entanglement. 

\begin{figure}[t]
  \begin{center}
  \includegraphics[width=0.99\columnwidth]{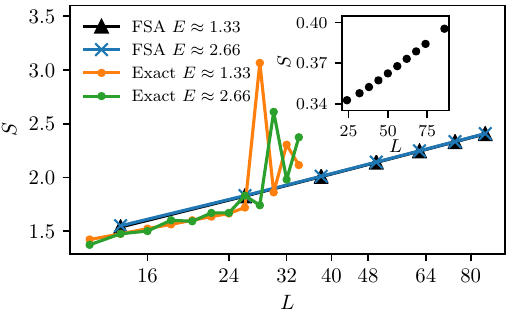}\\
  \caption{ \label{Fig:S_mps}
    Logarithmic scaling of entropy for two adjacent FSA eigenstates in the middle of the spectrum.  Black triangles correspond to the state at energy $E_{1} \approx 1.33$ and blue crosses to $E_{2} \approx 2.66$ (the two eigenstates have approximately the same  entanglement entropy with difference $\Delta S \sim 0.1\%$). The fit gives $S \propto 0.48\log(L)$.
     Green curve corresponds to the entropy of the exact special eigenstate at  $E_{1} \approx 1.33$.  
The non-monotonic behavior of entropy in this case is attributed to weak hybridization with volume-law entangled states at nearby energies.     
     The inset displays the entropy of the FSA ground state. The weak growth of entropy with $L$ is an artefact of the approximation, since the exact ground state is gapped and obeys area law for entropy. 
     }
  \end{center}
\end{figure}

Fig.~\ref{Fig:S_mps} shows the scaling of the bipartite entanglement entropy in special eigenstates  extracted using a MPS implementation of the FSA. Despite entropy being a non-local quantity, we again find good agreement between the FSA and exact diagonalization results for system sizes up to $L=30$. The logarithmic growth of entanglement entropy with system size $L$ suggests that special eigenstates cannot be efficiently represented by MPS in the thermodynamic limit. We note that jumps in the entropy growth of special eigenstates obtained via exact diagonalization, visible in Fig.~\ref{Fig:S_mps}, can be understood as accidental hybridization with eigenstates in the bulk. Because the majority of eigenstates carry an extensive amount of entropy (volume-law) in the middle of the spectrum, such jumps can be attributed to two-eigenstate resonances.

Notably, the FSA overestimates the entanglement entropy for $L\leq 30$. This trend is even more pronounced in the inset of Fig.~\ref{Fig:S_mps}, which shows the scaling of the ground state entanglement entropy obtained with the FSA. From exact diagonalization it is known that the system is gapped, and the bipartite entanglement entropy is expected to saturate at the value $S\approx 0.346$ in the thermodynamic limit. The observed slow linear growth is an indication of the error of the FSA and we expect it to reside within all eigenstates. However, since the prefactor of the observed linear growth is very small, for the system sizes considered, it is not visible in the logarithmic entropy growth of the highly excited states. 

We demonstrated that the FSA allows one to extract eigenenergies and other characteristics of special eigenstates. Overall, we find good agreement of these results with exact diagonalization. The fact that one can capture many-body eigenstates in the Hilbert space (that scales exponentially in $L$) with a basis of $L+1$ vectors is unexpected. As we discuss below, this reflects the relation between special eigenstates and unstable periodic orbits. The FSA provides a basis in the many-body Hilbert space that approximately captures the dynamics associated with the periodic orbit.

\subsection{FSA subspace as a basis for quantum scarred eigenstates \label{Sec:fsa-discussion}}

Until now we have discussed the phenomenology of special eigenstates. Several properties of these special eigenstates suggest their similarity to quantum scarred eigenstates in single-particle systems. In particular, special eigenstates are concentrated in parts of the Hilbert space~\cite{Turner2017}, have approximately equal energy spacing, and are easily accessible by preparing the system in certain product states. However, in order to put the relation between special eigenstates and quantum scars on a firm basis, one needs to generalize the notion of a classical trajectory to the many-body quantum case. 

One promising route for defining an analogue of a classical trajectory in the many-body case is provided by the time dependent variational principle (TDVP)~\cite{Haegeman},  which allows to systematically construct a manifold of low-entangled states that furnish an effective ``semiclassical" description of many-body dynamics.  In particular, Ref.~\onlinecite{Bernien2017} captured the revivals using bond dimension 2 variational ansatz for the  collective Rabi oscillations of atoms ${\bullet}{\circ} \leftrightarrow {\circ}{\bullet}$ between two different configurations of the unit cell. These oscillations can be viewed as a trajectory connecting $\ket{\mathbb{Z}_2}=\ket{{\bullet}{\circ}{\bullet}{\circ}\ldots}$ product state and its translated version, $\ket{\mathbb{Z}'_2} = \ket{{\circ}{\bullet}{\circ}{\bullet}\ldots}$.  In recent work~\cite{Harvardtobe}, the TDVP approach was  extended to a wider class of spin models, thus providing a general framework to explore quantum scarring in the dynamics of many-body systems by an analogy with the single-particle case.

While the TDVP approach allows one to extract some characteristics of special eigenstates (for example, the oscillation frequency approximately agrees with the energy separation between adjacent special eigenstates), it is not clear if such an approach can be used for describing the properties of individual scarred eigenstates, such as the entanglement structure and expectation values of local observables, and for understanding the finite-size behavior.
In this respect the FSA approach provides a description of the nearly periodic Hilbert-space trajectory that is complementary to TDVP. Above we demonstrated that the FSA constructs a basis of $L+1$ states directly in the many-body Hilbert space of a finite-size system. The special property of this basis is that it effectively captures the unitary evolution $e^{-iHt}\ket{\mathbb{Z}_2}$ that connects the N\'eel state and its translated version, $\ket{\mathbb{Z}'_2}$. Indeed, the dynamics in the many-body Hilbert space starting from $v_0=\ket{\mathbb{Z}_2}=\ket{{\bullet}{\circ}{\bullet}{\circ}\ldots}$ proceeds via an increasing number of flips that are generated by the forward-propagation part of the Hamiltonian, $H_+$. In particular, at the first step the dynamics generates one delocalized defect within $\ket{\mathbb{Z}_2}$ state. This state coincides with the second basis vector in the FSA basis, $v_1$, see Eq.~(\ref{Eq:one-defect}). Similarly, the $v_2$ vector from the FSA, with two defects on top of $\ket{\mathbb{Z}_2}$ initial state, corresponds to the second step of the trajectory. Hence, we conclude that the FSA captures the dominant subspace of the Hilbert space where the dynamics connecting  $\ket{\mathbb{Z}_2}$ and  $\ket{\mathbb{Z}'_2}$ states occurs. This is further supported by Fig.~\ref{fig:fid_fsa_sector} in Appendix. 

Finally, let us discuss other families of quantum scarred eigenstates within the language of TDVP and FSA. In addition to  special eigenstates with enhanced overlap with the $\ket{\mathbb{Z}_2}$ product state, we also observed  $\ket{\mathbb{Z}_3}$-generated band of special states in Fig.~\ref{fig:z2z3overlap}. This shows  that the PXP model has more than one periodic trajectory that leads to quantum scars. In particular, the $\ket{\mathbb{Z}_3}$-band of special eigenstates is related to oscillations between the three-site configuration, ${\bullet}{\circ} {\circ}$, and configurations ${\circ}{\bullet} {\circ}$, ${\circ} {\circ} {\bullet}$, obtained from it by translations. These oscillations can also be described within TDVP~\cite{Harvardtobe,wetobe}.  We note that it is also possible to describe the corresponding scarred eigenstates using the FSA scheme starting from $\ket{\mathbb{Z}_3}$ product state. Moreover, the first step of the FSA recurrence still remains exact. However, in this case the FSA recurrence is frustrated: starting from  ${\bullet}{\circ} {\circ}$ state, forward propagation  brings one into either of the translated configurations, ${\circ}{\bullet} {\circ}$ or ${\circ} {\circ} {\bullet}$. This fact may potentially explain the observation that the $\ket{\mathbb{Z}_3}$-band of special eigenstates is less separated from the continuum of other eigenstates in Fig.~\ref{fig:z2z3overlap}(a). In other words, the trajectory starting from $\ket{\mathbb{Z}_3}$ product state is more unstable, leading to weaker quantum many-body scars.  Nevertheless,  one still observes distinct periodic revivals of the many-body fidelity starting from $\ket{\mathbb{Z}_3}$ state, see Fig.~\ref{fig:fidelity}.

The observation of $\ket{\mathbb{Z}_2}$ and $\ket{\mathbb{Z}_3}$ trajectories and underlying sets of scarred eigenstates naively suggests that density wave states with larger periods will also give rise to scars. Clearly, Fig.~\ref{fig:fidelity} shows that this is not the case as already $\ket{\mathbb{Z}_4}$ product state features a complete absence of revivals. We attribute this to the fact that the FSA approximation ceases to be exact at the first step for $\ket{\mathbb{Z}_n}$ product state with $n\geq 4$. This signals that the underlying trajectories become too unstable to produce quantum scars. On the other hand,  product states that contain domain walls between different $\ket{\mathbb{Z}_2}$ and $\ket{\mathbb{Z}_3}$ patterns can potentially lead to another set of scarred eigenstates. We leave a detailed investigation of this issue to future work.

\section{Stability against perturbations \label{sec:pert}}

Our discussion so far has demonstrated that the FSA is helpful for developing intuition about  the structure of various families of quantum scarred eigenstates at high energy densities. Here we investigate the stability of $\ket{\mathbb{Z}_2}$ special eigenstates with respect to various perturbations of the Hamiltonian. We will rely on the FSA to develop intuition why special eigenstates are robust with respect to certain perturbations, or which kind of perturbations are most efficient in removing the periodic orbits. Furthermore, we discuss several deformations that bring the PXP model to exactly solvable points. Some of these deformations were found in Refs.~\onlinecite{FendleySachdev,LesanovskyMPS}.  Below we demonstrate that these perturbations are strong and remove the special eigenstates that are found in the PXP model.

\subsection{Physical perturbations\label{Sec:pert-phys}}

We consider the following perturbations of the PXP Hamiltonian,
\begin{subequations}
\label{Eq:HhopHmuHXXX}
\begin{eqnarray} \label{Eq:Hmu}
 \delta H_{0}&=& g_{0} \sum_j Q_j,
 \\
 \delta H_\text{nn}&=& g_\text{nn} \sum_j P_{j-1} (\sigma^+_j\sigma^-_{j+1}+\sigma^-_j\sigma^+_{j+1})P_{j+2}, \;\;\;\label{Eq:Hhop}
 \\
  \delta H_\text{nnn}&=& g_\text{nnn} \sum_j P_{j-1}X_j P_{j+1} X_{j+2}P_{j+3}.
  \label{Eq:HXXX}
\end{eqnarray}
\end{subequations}
The uniform chemical potential, $\delta H_{0}$, and constrained nearest neighbour hopping, $\delta H_{\text{nn}}$, result from the second order Schrieffer-Wolf transformation, $  H_\text{SW}^{(2)}
 = \epsilon^2 \mathcal{P}  H_1 \left( \mathcal{P}_{(1)} +(1/2) \mathcal{P}_{(2)}\right) H_1\mathcal{P}$, 
where $\mathcal{P}_{(1,2)}$ are projectors onto subspaces with one or two adjacent excitations.
The last perturbation, $\delta H_\text{nnn}$, physically corresponds to correlated next-nearest neighbour flips. We note that these perturbations lift the zero-mode degeneracy as they commute with both particle-hole symmetry and inversion operators. Thus, all perturbations in Eq.~(\ref{Eq:HhopHmuHXXX}) effectively remove the bipartite structure of the Hilbert space graph that is responsible for the appearance of zero modes, see Appendix~\ref{sec:zero}. 

\begin{figure}[t]
\begin{center}
\includegraphics[width=0.99\columnwidth]{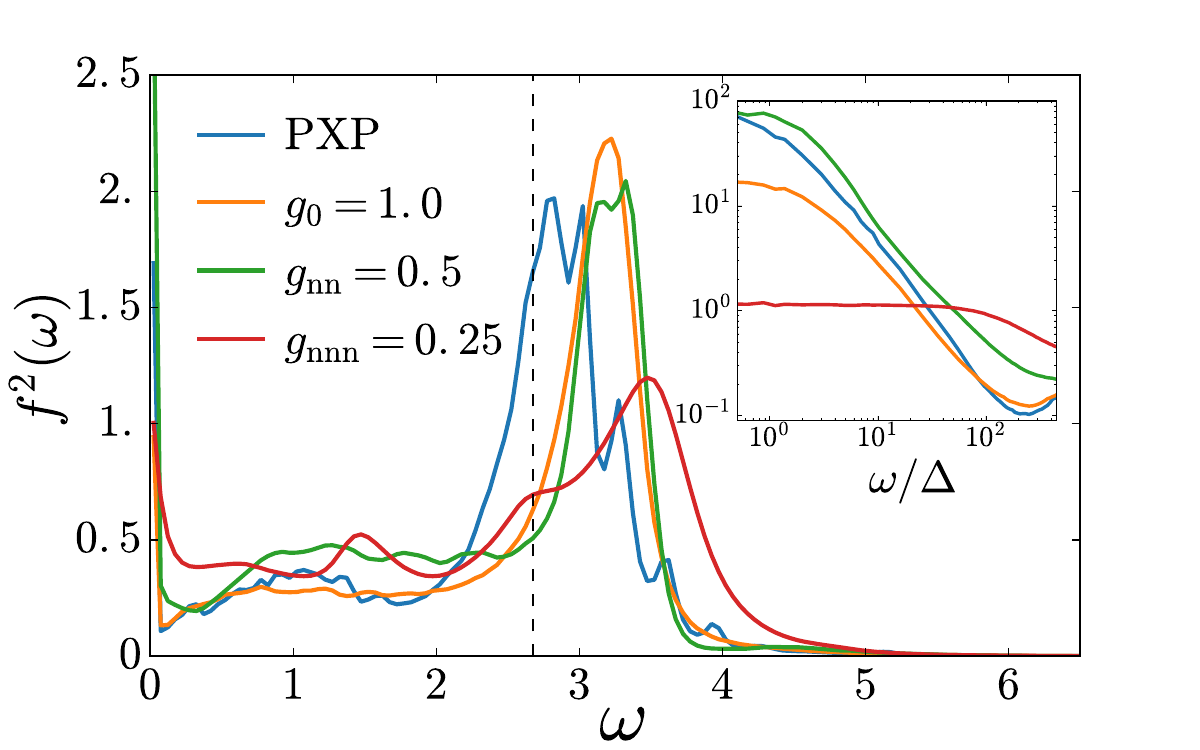}\\
\caption{ 
\label{Fig:f2-omega}
Peak in the energy dependence of matrix elements of the unperturbed PXP model at $\omega\approx 2.66$ softens and shifts upon the addition of perturbations.  Moreover, the inset shows that the plateau in $f^2(\omega)$  only slightly increases its size when the added perturbation is chemical potential or correlated hopping. In contrast, upon adding $\delta H_\text{nnn}$, the plateau increases to values of  $\omega/\Delta\leq 200$, fully consistent with the restoration of conventional thermalization. }
\end{center}
\end{figure}

As we discussed in Section~\ref{Sec:fsa-discussion}, the FSA allows to quantify the structure of the Hilbert-space orbit underlying quantum-scarred eigenstates. Hence, we use the intuition provided by the FSA to qualitatively understand sensitivity to different perturbations in Eq.~(\ref{Eq:HhopHmuHXXX}). In the case when the PXP model is perturbed by the uniform chemical potential, Eq.~(\ref{Eq:Hmu}), the FSA recurrence remains exact at the first and second steps.  However, $\delta H_{0}$ will introduce on-site energies in the FSA, making the diagonal of the tridiagonal matrix in Eq.~(\ref{eq:tmatrix}) non-zero. Hence, we expect that  $\delta H_{0}$ will change the frequency of oscillations and also contribute to their dephasing by removing the periodic energy spacing between special eigenstates. For the weak perturbation $g_{0}=0.2$,  we demonstrated almost no change in oscillations, see the Supplementary Material of Ref.~\onlinecite{Turner2017}. 
Moreover, in Fig.~\ref{Fig:f2-omega} we compare the structure of the spectral function in the PXP model  to that in the perturbed model with $g_0=1$. We observe that the peak in off-diagonal matrix elements $f^2(\omega)$ at $\omega\approx 2.66$ shifts to slightly larger frequencies, but still remains strongly pronounced. 

Next, we consider the case of the nearest neighbour hopping perturbation. This perturbation leaves the first step of the FSA exact, but introduces an error already at the second step. Yet, Fig.~\ref{Fig:f2-omega} shows that the peak in the spectral function associated with the separation between special eigenstates shifts while keeping the same magnitude when we add nearest neighbor hopping, $g_\text{nn}=0.5$. Note that the magnitude of the perturbation is chosen in such a way that it has comparable operator norm to the previously considered chemical potential with $g_0=1$. 

\begin{figure}
\begin{center}
\includegraphics[width=0.99\columnwidth]{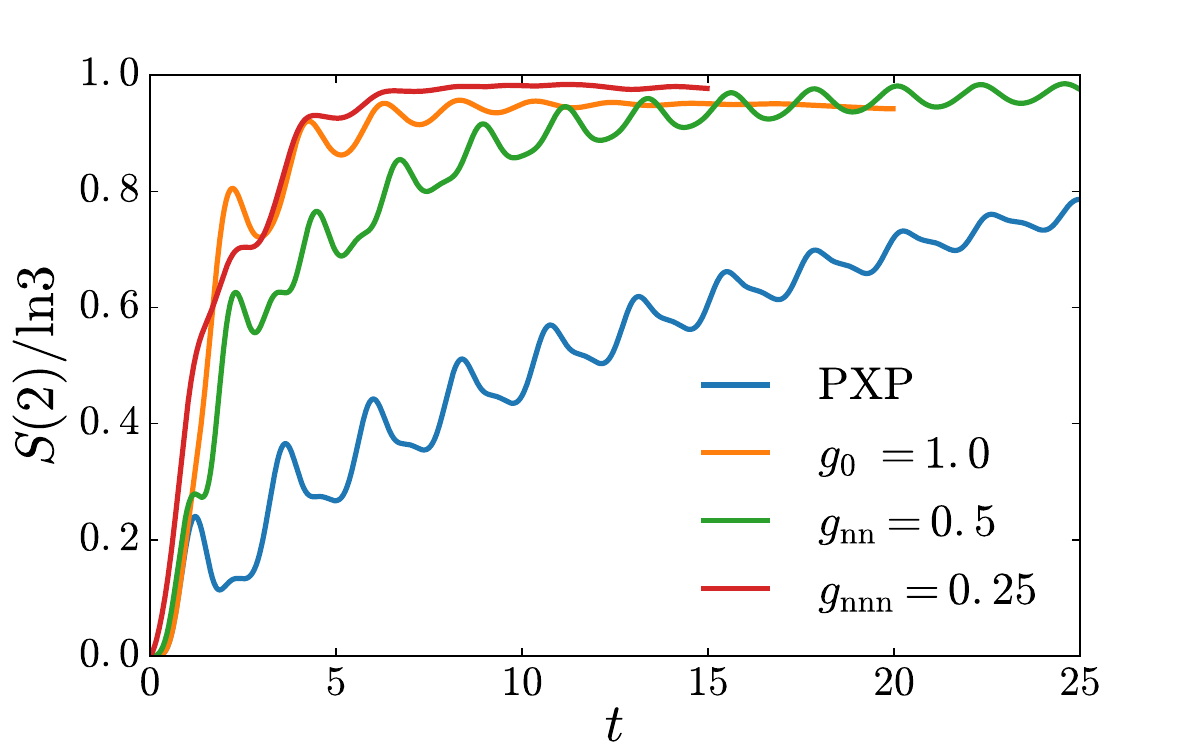}\\
\caption{ \label{Fig:f2-dynamics}
The dynamics of two-site entanglement in the quantum quench from $\ket{\mathbb{Z}_2}$ initial state is strongly influenced by perturbations to the PXP model. We note that nearest-neighbor hopping is the least effective in damping the oscillations. In contrast, for the perturbation $g_{\text{nnn}}=0.25$, when the bulk of eigenstates becomes fully thermal, the oscillations in entanglement and local observables are strongly damped. The entanglement is normalized by the maximal possible value for two sites, which is equal to $\ln 3$ due to the presence of a constraint. The data is obtained with iTEBD, the maximal evolution time is limited by the bond dimension $\chi=1200$. 
}
\end{center}
\end{figure}

Finally, we considered next nearest neighbor correlated flips, Eq.~(\ref{Eq:HXXX}), as a perturbation that introduces error even at the first step of the FSA approximation. Hence, we expect such a term to have the strongest effect of all three terms considered in Eq.~(\ref{Eq:HhopHmuHXXX}). Fully consistent with these expectations, we observe that the perturbation of magnitude $g_\text{nnn}=0.25$, which  has operator norm comparable to earlier perturbations,  suffices to significantly broaden the peak in the spectral function. In addition, we observe in Fig.~\ref{Fig:f2-dynamics} that this perturbation is the most efficient one in damping the oscillations of the local two-site entanglement after about three periods. We note that while the perturbations $g_\text{nnn}=0.25$ and $g_0=1$ both lead to the strongly enhanced growth of bipartite  entanglement with very similar slopes (not shown), the former is more efficient in damping the local oscillations.  

Above we observed that  perturbations $\delta H_0$ and $\delta H_\text{nn}$ are less effective in destroying the special eigenstates and the corresponding oscillations in the PXP model. The correlated flips $\delta H_\text{nnn}$ is most effective in destroying the oscillations. At the same time, we observe that the latter perturbation is most effective at removing the traces of ``slow" thermalization in the bulk of other eigenstates. In particular,  Fig.~\ref{fig:eth}(b) demonstrated that while the fluctuations in local observables decay exponentially with the system size, this decay is slower than expected from the ETH.  We checked that for $g_\text{nnn}=0.25$ the fluctuations of local observables decay as a square root of the Hilbert space dimension, $\overline{\Delta O^Z}\propto 1/\sqrt{{\cal D}_{0+}}$, fully consistent with the ETH expectations.  In addition, the inset of Fig.~\ref{Fig:f2-omega} shows that $\delta H_\text{nnn}$ perturbation corresponds to the best-developed plateau in the spectral function at small energy separations. 

Thus we conclude that the existence of well-defined special eigenstates on the one side, and anomalies in thermalization of the bulk of eigenstates  on the other hand, are related to each other. In other words, the existence of strongly scarred quantum many-body eigenstates and their ``protection'' from the bulk of other eigenstates is intertwined with slower thermalization of other eigenstates.

\subsection{Integrable deformations of PXP model \label{Sec:pert-int}}

After illustrating that perturbations that are effective in restoring ``canonical ETH'' thermalization also destroy the bands of special eigenstates, we discuss  deformations of the PXP model that make it exactly solvable. PXP Hamiltonian can be deformed to become integrable by a one-parameter family of deformations~\cite{FendleySachdev}. The set of integrable models includes the so-called ``golden chain'' Fibonacci anyonic model.~\cite{Feiguin07} In addition, there exist a one-parameter family of frustration-free Hamiltonians that includes the PXP term.~\cite{LesanovskyMPS} 

In particular, by adding the operator  
\begin{equation}\label{Eq:Int}
\delta H_v = - \sum_j {\cal P}\left[v Q_{j-1}  Q_{j+1} + \left(v^{-1} - v\right) Q_j\right]{\cal P}
\end{equation}
 with one free real parameter $v$ to the Hamiltonian of the PXP model, we obtain a Bethe-ansatz  solvable model~\cite{FendleySachdev}. For $v=1$ this perturbation amounts to the constant next-nearest neighbor interaction of Rydberg atoms. Another special point is $v=2^{-1/4}$, when the total norm of the operators in Eq.~(\ref{Eq:Int}) takes a minimal value. Moreover, this integrable line has a quantum critical/tricritical point at   $v_{3,2} = \mp (  (\sqrt{ 5} + 1)/2 )^ { 5/ 2}$,  respectively.~\cite{FendleySachdev,Sachdev18} 
 
 In the integrable models all eigenstates violate conventional ETH and can be described only via the generalized Gibbs ensemble that incorporates additional conserved quantities. At the same time, we explicitly checked for all these cases that the special eigenstates, found in unperturbed PXP model via overlap with $\mathbb{Z}_2$ product state, are either strongly perturbed or completely destroyed. Moreover, we did not observe any low-entanglement eigenstates at energy $E$ close to zero, unlike for the unperturbed PXP model, see Fig.~\ref{fig:entropyscatter}.

 Another family of perturbations with a free parameter~$z$, 
 \begin{equation}\label{Eq:FF}
\delta H_z=\sum_j (z P_{j-1}P_{j}P_{j+1}+z^{-1} P_{j-1}Q_jP_{j+1}),
\end{equation}
brings the PXP model into a frustration-free Hamiltonian, which allows for the exact solution of its ground state~\cite{LesanovskyMPS} for any real value of $z$. When $z\to 0$ ($z\to\infty$), the first (second) term dominates, and this perturbation always has finite magnitude $O(1)$ for any values of $z$. Analytically minimizing the energy of the PXP model Eq.~(\ref{eq:H_PXP}) using a frustration-free MPS ansatz of Ref.~\onlinecite{LesanovskyMPS} results in the value $z_\text{PXP}=(\sqrt{5}+1)^{1/2}/(2\sqrt{2}) \approx 0.636$. Such an  approximation reproduces local observables of the exact ground state, such as energy density, with high precision. Perturbing the PXP model using $\delta H_z$, we find that fidelity oscillations  always decay faster compared to the unperturbed PXP model. For $z \sim z_\text{PXP}$ we find the slowest decay for all $z$ so that the perturbation $\delta H_{z_\text{PXP}}$ has weakest effect. Damping of oscillations increases with increasing $|z- z_\text{PXP}|$.
 
In summary, we observed that deforming the PXP model to nearby solvable points does not improve the robustness of quantum scarred eigenstates. Instead, such deformations either strongly perturb these states or lead to their complete disappearance. This result suggests that it is the unperturbed PXP model that should be viewed as a parent model for the quantum-scarred eigenstates, and despite proximity of several integrable points to this model,  Bethe-ansatz integrability cannot be used to explain this behavior.

\section{Discussion and outlook \label{sec:disc}}

In this paper we studied the eigenstate and dynamical properties of the PXP model, which describes a chain of Rydberg atoms realized in recent experiments~\cite{Bernien2017}. We found that the majority of the eigenstates of this model thermalize more slowly compared to other microscopic models that are usually used to test the ETH~\cite{Alessiorev}. 
 On the one hand, the origin of such behavior may be related to the constraints in the PXP model, which make the Hamiltonian sparse in the Hilbert space. On the other hand, conventional ETH was shown to hold for other kinematically constrained models~\cite{Chandran16}. Hence, we speculate that anomalies in the thermalization in the PXP model may be related to the existence of quantum scarred eigenstates.

These quantum scarred eigenstates, identified in Ref.~\onlinecite{Turner2017}, strongly violate the ETH. In particular, these special eigenstates stand out due to their anomalous expectation values of local observables, as well as their much smaller entanglement entropy compared to thermal eigenstates with similar energies. Besides the ETH violation,  special eigenstates are characterized by their large overlaps with charge density wave states $\ket{\mathbb{Z}_2}$ and $\ket{\mathbb{Z}_3}$. The energies of these special eigenstates are (approximately)  multiples of the same fundamental frequency. Consequently, as discussed in Ref.~\onlinecite{Turner2017} and in this paper, these eigenstates play a key role in the experimentally observed many-body revivals in the quantum quench setup~\cite{Bernien2017}. We also predicted the existence of revivals for the case when the system is initialized in $\ket{\mathbb{Z}_3}$ product state, with some additional features compared to the $\ket{\mathbb{Z}_2}$ case, and we identified the corresponding family of special eigenstates.

The phenomenology of the special eigenstates described above allowed us to draw parallels with the ubiquitous phenomenon of quantum scarring, thus lending support to the  term {\it quantum many-body scars}. In the case of a single particle in a chaotic billiard, a single unstable periodic classical orbit leads to a set of scarred eigenstates~\cite{Heller84}. These eigenstates have their wave functions localized in the vicinity of their parent trajectory, and can be efficiently prepared by initializing the wave packet near the classical orbit. Moreover,  in chaotic billiards one usually finds  more than one periodic trajectory that gives rise to quantum scars. The classical trajectories which are less unstable give rise to more localized wave functions, corresponding to stronger scarring. Finally, one also expects some degree of stability of quantum scars to perturbing the system, unless the perturbation destroys the periodic trajectory. 

Similarly, in the PXP model, we observed a set of $L+1$ special eigenstates that are well described within the FSA basis, whose dimension scales linearly with the  system size. This suggests that special eigenstates are concentrated in a small part of the Hilbert space, analogous to the case of a chaotic billiard. The system effectively accesses these eigenstates when prepared in the initial $\ket{\mathbb{Z}_2}$ state or its translated partner, $\ket{\mathbb{Z}'_2}$. Additionally, in this work we reported a second family of scarred eigenstates arising from $\ket{\mathbb{Z}_3}$ density wave product state. This second family of eigenstates has larger entanglement, suggesting that the underlying orbit is less stable. The enhanced stability of $\ket{\mathbb{Z}_2}$ special eigenstates compared to their $\ket{\mathbb{Z}_3}$ counterparts can also be understood within the FSA.

Finally, we demonstrated the stability of the many-body revivals with respect to perturbations of the PXP model. We confirmed that perturbations which do not introduce any immediate errors in the FSA approach are less effective in destroying the bands of special eigenstates and the revivals. We also identified a perturbation that quickly removes the non-ergodic scarred eigenstates, restoring the ETH for all states. 
In addition, we also considered several deformations of the PXP model that bring it to solvable points. Although the PXP model can be deformed into Bethe-ansatz integrable models which do not follow the ETH, the characteristic many-body revivals from simple product states do not persist in these integrable models. Thus, the proximity of those integrable lines is likely unrelated to the weak ergodicity breaking in the PXP model.

While our study sheds new light onto the structure and stability of quantum many-body scars, many interesting questions remain open. We used the FSA approximation throughout this paper and  provided a simple estimate of the incurred errors in Appendix~\ref{App:error}. However, quantifying the final error in the  FSA remains an open problem. Furthermore, it would be highly desirable to identify a parameter that governs the stability of quantum scars in the generic case. 
Better understanding of the errors in the FSA would allow to obtain more rigorous understanding of quantum-scarred eigenstates in the thermodynamic limit. While the FSA suggests their persistence, numerical studies have revealed an onset of accidental hybridizations between scarred eigenstates and the thermalizing bulk of eigenstates. These hybridizations resulted in irregular behavior of entanglement entropy for larger system sizes $L\geq 34$, despite the energies of special eigenstates still following accurate finite-size scaling. Generally,  one may expect special eigenstates to get ``dissolved" in the bulk in the thermodynamic limit.  Nevertheless, there will be signatures remaining in the properties of local operators and dynamics at short and intermediate time scales.  In particular, the structure of the spectral function reported in this work is expected to be robust in the thermodynamic limit. For instance, the unusual peaks in the off-diagonal matrix elements at the energy difference of order one, shown in Fig.~\ref{fig:eth}(c) and Fig.~\ref{Fig:f2-omega}, are converged with the system size. Revealing and identifying other experimentally observable signatures remains an interesting problem.

More broadly, it would be desirable to understand whether there are wider classes of models that display quantum scars. On the one hand, the FSA suggests that the constraint present in the PXP model plays a crucial role in protecting and enabling such behavior. Therefore,  it would be natural to search for other types of (constrained) Hilbert spaces and models with similar behavior.  For example, we note that the model in Eq.~(\ref{eq:H_PXP}) is related to a class of models that represent interactions between fundamental excitations in topological phases of matter in two dimensions~\cite{Feiguin07,  Lesanovsky2012, Lindner2012, Glaetzle2014, Vasseur2015, Chandran16, Lan2017, Lan2017_2}. A wide class of such phases are the fractional quantum Hall states, in which electrons fractionalize into Abelian or non-Abelian anyons. In particular, in a $\nu=12/5$ fractional quantum Hall state, the fundamental excitation is a Fibonacci anyon $\tau$~\cite{ReadRezayi}. The rules of anyon fusion place a formally similar constraint to the allowed number of anyons as our constraint on the allowed excitations in the Rydberg atom chain. Thus, it would be interesting to explore analogous models (in the context of cold atomic gases or trapped ions) for different types of anyon models, and investigate the occurrence and stability of quantum scars in them.

Finally, the issues discussed above naturally connect to questions about \emph{practical} uses of quantum many-body scars and their dynamical signatures. Preparing the system in a superposition of quantum-scarred eigenstates effectively shields it from thermal relaxation for much longer times. Hence,  a better understanding of fundamental properties of quantum many-body scars, their stability and tunability may be of potential use in experiments studying dynamics of non-equilibrium many-body quantum systems.

\emph{Note added.---} Very recently,  Ref.~\onlinecite{BernevigEnt} has analytically constructed a set of non-thermalizing eigenstates in the AKLT model with logarithmic scaling of the entanglement entropy. As discussed in Ref.~\onlinecite{BernevigEnt}, the existence of such eigenstates is suggestive of the presence of quantum scars in the AKLT model.

Furthermore, during the completion of this manuscript, we became aware of two related works on the PXP model~\cite{Harvardtobe, Khemani2018}. In Ref.~\onlinecite{Harvardtobe},
a generalization of the TDVP approach for various spin models and a connection with periodic orbits has been developed. In a different direction,  Ref.~\onlinecite{Khemani2018} has argued that special properties of the PXP model result from a ``proximate integrable point", to which the model can be driven by applying a particular perturbation.

\section*{Acknowledgments}

We thank Paul Fendley, Misha Lukin, Hannes Pichler, Marcos Rigol, Harry Levine, and Wen Wei Ho for illuminating discussions.  C.J.T. and Z.P. acknowledge support by EPSRC grants EP/P009409/1 and EP/M50807X/1, and the Royal Society Research Grant RG160635. D.A. acknowledges support by the Swiss National Science Foundation. A.M. and M.S. acknowledge support provided by  J. Kiss and A. Schl\"ogl from the HPC Scientific Service Unit of IST Austria. Statement of compliance with EPSRC policy framework on research data: This publication is theoretical work that does not require supporting research data.

\appendix

\section{Zero-energy states}\label{sec:zero}
 
In the main text it was mentioned that one of the special features of the PXP model is the existence of an exponentially large number of states which are annihilated by the PXP Hamiltonian in Eq.~(\ref{eq:H_PXP}).
In Ref.~\onlinecite{Turner2017} (see also Ref.~\onlinecite{Iadecola2018}) it was shown that the degeneracy of this zero-energy subspace, $\mathcal{Z}_L$, grows with system size according to a Fibonacci number $F$. More precisely, for open boundaries, depending on whether the system size $L$ is even or odd, we have
\begin{align}
  \mathcal{Z}_{2n} &= F_{n+1}, & \mathcal{Z}_{2n+1} &= F_n.
\end{align}
For periodic boundaries in the zero-momentum sector
\begin{align}
  \mathcal{Z}_{2n}^{(0)} &= F_{n-1}, & \mathcal{Z}_{2n+1}^{(0)} &= F_{n-1},
\end{align}
whilst in the $\pi$-momentum sector instead
\begin{align}
  \mathcal{Z}_{2n}^{(\pi)} &= F_{n-2}\text{.}
\end{align}
We note that there are zero energy levels in other symmetry sectors, but they are fewer in number and they will not be explicitly considered here.

In this Appendix, we formally derive the above counting for both OBC and PBC (in the zero momentum sector).
The key to this is the particle-hole symmetry, generated by the operator
\begin{equation}
  \mathcal{C} = \prod_i Z_i
\end{equation}
which anticommutes with the PXP Hamiltonian, $ {\cal C}H=-H {\cal C}$. Each eigenstate $|\psi\rangle$ with energy $E\neq 0$ therefore has a partner $ {\cal C}|\psi\rangle$ with energy $-E$.
The graph has a bipartite structure with vertex subsets that are even and odd in the number of excitations, which are measured by $\mathcal{C}$.
It is well known that the difference in dimensions of these two subspaces lower bounds the number of zero energy states~\cite{Sutherland86,Inui}.
However, applying this idea directly, gives us the difference between sectors with an even and odd number of excitations to be at most one, which is not a useful lower bound.
Missing from this analysis is consideration of the inversion symmetry $I$,
\begin{eqnarray}\label{eq:inversion}
I:\qquad j \mapsto L - j +  1,
\end{eqnarray}
which commutes with $\mathcal{C}$ and hence in its symmetry sectors the bipartite structure is preserved.
The combined action of these two symmetries will be shown to provide a tight bound for the number of zero energy states. We note that the exponentially-large number of zero-energy states is
an interesting feature of the PXP model because energy $E=0$ corresponds to the middle of the many-body spectrum.
By contrast, in 2D models endowed with supersymmetry, exponentially many zero-energy states can occur in the ground state manifold~\cite{FendleySchoutens}.

Curiously, the zero-mode degeneracy is robust for even $L$ to perturbation by the staggered magnetic field $\sum_j (-1)^j Z_j$, yielding for open boundary conditions
\begin{align}
  \mathcal{Z}_{2n} &= F_{n+1}, & \mathcal{Z}_{2n+1} &= 0\text{.}
\end{align}
For periodic boundaries, the staggered field explicitly breaks translation symmetry to a $\mathbb{Z}_{L/2}$ subgroup, thereby combining the zero- and $\pi$-momentum sectors
\begin{align}
  \mathcal{Z}_{2n}^{(0,\pi)} = F_{n-1} + F_{n-2}\text{.}
\end{align}
From the analysis below, it follows that zero modes are generally present if the Hamiltonian anticommutes with the product of particle-hole symmetry $\mathcal{C}$ and inversion $I$. Staggered field is a special case of this as it commutes
with particle-hole symmetry $\mathcal{C}$ and anticommutes with inversion $I$.

\subsection{Open chain}

Our Hilbert space $\mathcal{H} = \mathbb{C}[V]$, where $V$ is the vertex set, decomposes into subspaces containing states with even and odd numbers of excitations.
Those are measured by $\mathcal{C}$, and will be denoted by subscripts $e$ and $o$.
Each of these subspaces further decomposes into orbits under the action of the inversion operator $I$.
The orbits of $I$ are either one- or two-dimensional.
Denote the subspaces spanned by even invariant elements $\mathcal{K}_e$ and odd invariant elements $\mathcal{K}_o$.
Each two-element orbits contains one inversion-even irreducible representation and one inversion-odd irreducible representation. We denote these as $\mathcal{M}_{o/e}^{\pm}$, where $+$ means reflection  even and $-$ means reflection odd. In what follows we will use Latin letters for the dimension of the vector spaces labelled by the corresponding script letter.

In each of the $I$-sectors there is a lower bound on the number of zero energy states given by the difference between the dimensions of the subspaces of even and odd numbers of excitations.
Putting this together,
\begin{eqnarray}\label{eq:zed}
  \mathcal{Z}_L
  &\ge& \left||\mathcal{M}_e^+ \oplus \mathcal{K}_e | - |\mathcal{M}_o^+ \oplus \mathcal{K}_o| \right| + \left||\mathcal{M}_e^-| - |\mathcal{M}_o^-|\right| \nonumber \\
  &=& |M_e + K_e - M_o - K_o| + |M_e - M_o| \nonumber \\
  &\ge& |K_e - K_o|,
\end{eqnarray}
where we have used the triangle inequality.
All that remains is to calculate the vector space dimensions $K_e$ and $K_o$.

Before deriving general expressions for $K_e$ and $K_o$, we present a simple example to illustrate the above. For a chain of size $L=4$ with OBC, the Hilbert space contains 8 states in total, four of which are even in the number of excitations,
\begin{align}
  &&&&{\bullet}{\circ}{\bullet}{\circ},&&{\bullet}{\circ}{\circ}{\bullet},&&{\circ}{\bullet}{\circ}{\bullet},&&{\circ}{\circ}{\circ}{\circ},&&&&
\end{align}
and four odd ones,
\begin{align}
  &&&&{\bullet}{\circ}{\circ}{\circ},&&{\circ}{\bullet}{\circ}{\circ},&&{\circ}{\circ}{\bullet}{\circ},&&{\circ}{\circ}{\circ}{\bullet}.&&&&
\end{align}
Two of these (${\bullet}{\circ}{\circ}{\bullet}$,${\circ}{\circ}{\circ}{\circ}$) are invariant under $I$, while the rest can be organized into two-dimensional orbits. Thus, 
\begin{eqnarray}
\mathcal{M}_e^+ &=& \Big\{ {\bullet}{\circ}{\bullet}{\circ} + {\circ}{\bullet}{\circ}{\bullet}  \Big\}, \notag \\
\mathcal{M}_o^+ &=& \Big\{ {\bullet}{\circ}{\circ}{\circ} + {\circ}{\circ}{\circ}{\bullet},\;\;\;{\circ}{\bullet}{\circ}{\circ} + {\circ}{\circ}{\bullet}{\circ} \Big\}, \notag \\
\mathcal{K}_e &=& \Big\{ {\bullet}{\circ}{\circ}{\bullet},\;\;\;{\circ}{\circ}{\circ}{\circ} \Big\}, \notag \\
\mathcal{K}_o &=& \Big\{ \Big\}, \notag \\
\mathcal{M}_e^- &=& \Big\{ {\bullet}{\circ}{\bullet}{\circ} - {\circ}{\bullet}{\circ}{\bullet} \Big\}, \notag \\
\mathcal{M}_o^- &=& \Big\{ {\bullet}{\circ}{\circ}{\circ} - {\circ}{\circ}{\circ}{\bullet},\;\;\;{\circ}{\bullet}{\circ}{\circ} - {\circ}{\circ}{\bullet}{\circ} \Big\}. \notag
\end{eqnarray}
Plugging into Eq.~(\ref{eq:zed}), we find $\mathcal{Z}_{L=4} \geq |3-2| + |2-1| = 2$, which indeed agrees with the exact result $\mathcal{Z}_{L=4}=2$.

Next, we consider the case of general $L$.
For any configuration $A_{n-1}$ on the open chain of length $n-1$, there is a corresponding invariant element on the length $L=2n$ chain (with $\mathcal{C}=+1$) given by 
\begin{align}
  A_{n-1}{\circ}{\circ}A_{n-1}^T &\in \mathcal{K}_e,
\end{align}
where $A_n^T$ is the spatially reversed pattern of $A_n$.
Every element of $\mathcal{K}_e$ is of this form because the central two sites cannot contain excitations as they would then be adjacent.
These configurations are in one-to-one correspondence and therefore $K_e = F_{n+1}$ and $K_o=0$, giving $\mathcal{Z}_{2n} = F_{n+1}$ for open chains with even length.

Similarly, for $L = 2n + 1$ odd, there is again a one-to-one correspondence between invariant configurations of fixed excitation parity and configurations of smaller open chains.
In particular, the invariant configurations can be constructed for the two sectors as follows
\begin{align}
  \mathcal{C}&=+1 &:&& A_{n}{\circ}A_{n}^T &\in \mathcal{K}_e, \\
  \mathcal{C}&=-1 &:&& A_{n-1}{\circ}{\bullet}{\circ}A_{n-1}^T &\in \mathcal{K}_o.
\end{align}
This reveals that $K_e = F_{n+2}$ and $K_o = F_{n+1}$ for odd length open chains, which altogether gives $\mathcal{Z}_{2n+1}=F_n$.

\subsection{Open chain with alternating field}

In this subsection we generalize the lower bound on the zero-energy degeneracy to the case of open chains in the presence of a staggered field $S \equiv \sum_j (-1)^j Z_j$.
First, notice that
\begin{equation}
  SI = \begin{cases}
     - IS, & \text{ if $L$ even,} \\
     + IS, & \text{ if $L$ odd,}
  \end{cases}
\end{equation}
i.e., $[S,I]=0$ if $L$ is odd and $\{S,I\}=0$ if $L$ is even, where $I$ is the inversion symmetry in Eq.~(\ref{eq:inversion}).

Assume $L$ even and let $X$ be our unperturbed Hamiltonian. We can partition the Hilbert space in the following manner,
\begin{align}
  X + S \simeq \,\,
  \begin{blockarray}{ccccccc}
    \mathcal{M}_o^{+} & \mathcal{M}_e^{-} & \mathcal{M}_e^{+} & \mathcal{K}_e & \mathcal{M}_o^{-} \\
    \begin{block}{(cc|ccc)cc}
       &   & X & X & S & \,\,\mathcal{M}_o^{+} & \multirow{2}{*}{\bigg\}$\mathcal{H}_\downarrow$}\\
       &   & S &   & X & \,\,\mathcal{M}_e^{-}\\
     \cline{1-5}
     X & S &   &   &   & \,\,\mathcal{M}_e^{+} & \multirow{3}{*}{\Bigg\}$\mathcal{H}_\uparrow$} \\
     X &   &   &   &   & \,\,\mathcal{K}_e \\
     S & X &   &   &   & \,\,\mathcal{M}_o^{-} \\
    \end{block}
  \end{blockarray}.
  \label{eq:alt}
\end{align}
In this block diagram, an $X$ or $S$ denotes a non-zero block of $X + S$ according to the partitioning of the Hilbert space in terms of orbits of $\mathcal{C}$ and $I$ that was discussed in the previous section.
Note that $\mathcal{K}_o$ does not appear because it's trivial for $L$ even.

The block structure arises from symmetry considerations.
Since $X$ anticommutes with $\mathcal{C}$ and commutes with $I$, its non-zero matrix elements couple sectors which have different excitation parities and the same inversion parities.
Opposite to this, $S$ commutes with $\mathcal{C}$ and anticommutes with $I$. Accordingly, its non-zero matrix elements couple sectors with the same excitation parities but different inversion parities.

The blocks in Eq.~(\ref{eq:alt}) have been judiciously arranged to reveal a bipartite structure between subspaces $\mathcal{H}_\uparrow$ and $\mathcal{H}_\downarrow$.
This occurs because $X + S$ anticommutes with $\mathcal{C}I$.
From this bipartite structure we get a bound on the zero-energy degeneracy,
\begin{align}
  \mathcal{Z}_{2n} \ge | H_\uparrow - H_\downarrow | &= | M_e + K_e + M_o - M_o - M_e | \nonumber \\
  &= K_e = F_{n+1},
\end{align}
which is tight, i.e., matches the exact number of zero-energy states in the presence of staggered field.

Interestingly, for odd $L$ the situation is completely different. Now the $S$-blocks go along the diagonal, which removes the bipartite structure. Thus, we do not expect any zero-energy states in this case, which is indeed confirmed by exact calculation.

Beyond this example we can see that the bipartite block-structure between $\mathcal{H}_\uparrow$ and $\mathcal{H}_\downarrow$ is undisturbed by the addition of terms to the Hamiltonian which anticommute with $\mathcal{C}I$ and do not violate the adjacency constraint.
This is because $\mathcal{H}_\uparrow$ and $\mathcal{H}_\downarrow$ are the subspace of $\mathcal{C}I = +1$ and $-1$ respectively since $\mathcal{C}$ measures the number parity of excitations.

\subsection{Periodic chain}

For a periodic chain, in addition to $\mathcal{C}$ and $I$, we also need to consider the cyclic translation generator $\sigma$,
\begin{eqnarray}
  \sigma: \qquad j \mapsto (j + 1)\mod{L}.
\end{eqnarray}
Every translation orbit of a periodic chain of length $L$ that is left invariant by $I$ contains at least one element invariant under $I$ or $I\sigma$, i.e., under either a site or a bond inversion.

First, take $L=2n+1$ odd, and consider the orbits which are excitation odd;
the inversion invariant orbits must contain at least one element of the form
\begin{equation}
  \begin{tikzpicture}
    \draw (360/4+0*360/17: 0.4cm) node{${\bullet}$};
    \draw (360/4+1*360/17: 0.4cm) node{${\circ}$};
    \draw (360/4-1*360/17: 0.4cm) node{${\circ}$};
    \draw (360/4+8*360/17: 0.4cm) node{${\circ}$};
    \draw (360/4+9*360/17: 0.4cm) node{${\circ}$};

    \draw (360/4+4*360/17: 0.4cm) node{$A$};
    \draw (360/4-4*360/17: 0.4cm) node{$A^T$};

    \draw (0,-0.6cm) -- (0,+0.6cm);
    \draw[dashed] (360/4: 0.5cm) -- (-3*360/4: 0.5cm);
  \end{tikzpicture}
\end{equation}
This diagram depicts a configuration wrapped around a ring.
As before, $A$ denotes an arbitrary pattern (which connects to ${\circ}{\circ}$ and ${\circ}{\bullet}{\circ}$ on its two ends), while $A^T$ is the spatially reversed pattern of $A$.
Suppose then that this configuration is non-unique with
\begin{equation}
  \begin{tikzpicture}[baseline=-0.1cm]
    \draw (360/4+0*360/17: 0.4cm) node{${\bullet}$};
    \draw (360/4+1*360/17: 0.4cm) node{${\circ}$};
    \draw (360/4-1*360/17: 0.4cm) node{${\circ}$};
    \draw (360/4+8*360/17: 0.4cm) node{${\circ}$};
    \draw (360/4+9*360/17: 0.4cm) node{${\circ}$};

    \draw (360/4+4*360/17: 0.4cm) node{$A$};
    \draw (360/4-4*360/17: 0.4cm) node{$A^T$};

    \draw (0,-0.6cm) -- (0,+0.6cm);
    \draw[dashed] (360/4: 0.5cm) -- (-3*360/4: 0.5cm);
  \end{tikzpicture}
  =
  \sigma^j
  \begin{tikzpicture}[baseline=-0.1cm]
    \draw (360/4+0*360/17: 0.4cm) node{${\bullet}$};
    \draw (360/4+1*360/17: 0.4cm) node{${\circ}$};
    \draw (360/4-1*360/17: 0.4cm) node{${\circ}$};
    \draw (360/4+8*360/17: 0.4cm) node{${\circ}$};
    \draw (360/4+9*360/17: 0.4cm) node{${\circ}$};

    \draw (360/4+4*360/17: 0.4cm) node{$B$};
    \draw (360/4-4*360/17: 0.4cm) node{$B^T$};

    \draw (0,-0.6cm) -- (0,+0.6cm);
    \draw[dashed] (360/4: 0.5cm) -- (-3*360/4: 0.5cm);
  \end{tikzpicture}
\end{equation}
for some $j$ and $B \neq A$.
Take these two mirror planes and generate the full set of mirror planes. The invariant element then takes the form
\begin{equation}
  \begin{tikzpicture}
    \draw (360/4+0*360/36: 0.8cm) node{${\bullet}$};
    \draw (360/4+1*360/36: 0.8cm) node{${\circ}$};
    \draw (360/4-1*360/36: 0.8cm) node{${\circ}$};
    \draw (360/4+17.5*360/36: 0.8cm) node{${\circ}$};
    \draw (360/4+18.5*360/36: 0.8cm) node{${\circ}$};
    \draw (0,-0.9cm) -- (0,+0.9cm);

    \draw (360/4+12*360/36: 0.8cm) node{${\bullet}$};
    \draw (360/4+13*360/36: 0.8cm) node{${\circ}$};
    \draw (360/4+11*360/36: 0.8cm) node{${\circ}$};
    \draw (360/4-5.5*360/36: 0.8cm) node{${\circ}$};
    \draw (360/4-6.5*360/36: 0.8cm) node{${\circ}$};
    \draw[rotate=360/3] (0,-0.9cm) -- (0,+0.9cm);

    \draw (360/4-12*360/36: 0.8cm) node{${\bullet}$};
    \draw (360/4-13*360/36: 0.8cm) node{${\circ}$};
    \draw (360/4-11*360/36: 0.8cm) node{${\circ}$};
    \draw (360/4+5.5*360/36: 0.8cm) node{${\circ}$};
    \draw (360/4+6.5*360/36: 0.8cm) node{${\circ}$};
    \draw[rotate=-360/3] (0,-0.9cm) -- (0,+0.9cm);

    \draw (360/4+3*360/36: 0.8cm) node{$Y$};
    \draw (360/4+8.5*360/36: 0.75cm) node{$Y^T$};
    \draw (360/4+15*360/36: 0.8cm) node{$Y$};
    \draw (360/4-15*360/36: 0.85cm) node{$Y^T$};
    \draw (360/4-8.5*360/36: 0.8cm) node{$Y$};
    \draw (360/4-3.5*360/36: 0.9cm) node{$Y^T$};
\end{tikzpicture}
\end{equation}
from which it follows that both $A$ and $B$ must have the form ${\circ}Y{\circ}{\circ}Y^T{\circ}{\bullet}{\circ}Y$, thus they are equal.
This demonstrates that the inversion-invariant element in each inversion-invariant translation orbit is unique.
This one-to-one correspondence with the allowed configurations of open chains of length $n-2$ provides $K_o^{(0)} = F_n$.
If the orbit is instead excitation even, the diagram is instead
\begin{equation}
  \begin{tikzpicture}
    \draw (360/4+0*360/17: 0.4cm) node{${\circ}$};
    \draw (360/4+8*360/17: 0.4cm) node{${\circ}$};
    \draw (360/4+9*360/17: 0.4cm) node{${\circ}$};

    \draw (360/4+4*360/17: 0.4cm) node{$A$};
    \draw (360/4-4*360/17: 0.4cm) node{$A^T$};

    \draw (0,-0.6cm) -- (0,+0.6cm);
  \end{tikzpicture}
\end{equation}
and the same reasoning can be applied to find $K_e^{(0)} = F_{n+1}$.

Now take $L=2n$ even and consider the excitation-odd inversion-invariant translation orbits; these must contain an inversion-invariant element of the form
\begin{equation}
  \begin{tikzpicture}[baseline=-0.1cm]
    \draw (360/4+0*360/16: 0.4cm) node{${\bullet}$};
    \draw (360/4+1*360/16: 0.4cm) node{${\circ}$};
    \draw (360/4-1*360/16: 0.4cm) node{${\circ}$};
    \draw (360/4+8*360/16: 0.4cm) node{${\circ}$};

    \draw (360/4+4*360/16: 0.4cm) node{$A$};
    \draw (360/4-4*360/16: 0.4cm) node{$A^T$};

    \draw (0,-0.6cm) -- (0,+0.6cm);
  \end{tikzpicture}
  \text{.}
\end{equation}
The previous reasoning can again be applied to find $K_o^{(0)} = F_n$.

The final case is that of $L$ even and excitation-even orbits.
An elementary method like the previous is more involved for this case because the invariant elements are no longer unique.
We instead observe that our lower bound for $\mathcal{Z}$ is $K - 2 K_o$ where $K = K_e + K_o$ is the number of invariant elements irrespective of excitation parity.
The number $K$ can be found as $K=2(M + K) - (2M + K)$, where $M = M_e + M_o$ is the number of two-element orbits irrespective of excitation parity.
Note $2M + K$ is the number of translation orbits and $M+K$ is the number of orbits of the combined dihedral symmetry of translation and inversion symmetry.
These are integer sequences of system size and can be found in the OEIS as A000358 and A129526 respectively~\cite{Ashrafi:2016gb,OEIS}.
The ordinary generating functions of these sequences are known to be
\begin{align}
  2M + K &= [x^N]\sum_{k\ge1}\frac{\phi(k)}{k}\ln\frac{1}{1-x^k(1+x^k)} \\
  M + K &= 
\begin{aligned}[t]
  [x^N]\bigg(\frac{1}{2}\sum_{k\ge1}\frac{\phi(k)}{k}\ln\frac{1}{1-x^k(1+x^k)} \\[-1.2ex]%
  - \frac{1}{2}\frac{(1+x)(1+x^2)}{x^4+x^2-1}\bigg)\text{.}
\end{aligned}
\end{align}
where $\phi(k)$ is the Euler totient function, i.e., the number of positive integers up to $k$ that are relatively prime to $k$. Taking the appropriate linear combination of the generating functions
\begin{align}
  2(M+K)-(2M+K) &= -[x^N]\frac{(1+x)(1+x^2)}{x^4+x^2-1},
\end{align}
we recognize on the right hand side the generating function of the Fibonacci sequence, hence
\begin{align}
  2(M+K)-(2M+K) = F_{\left\lfloor N/2 \right\rfloor+2}.
\end{align}
From here we arrive at the desired result $K_e^{(0)} = F_{n+1}$, which completes the derivation of the zero energy degeneracy for even $L$ in the zero momentum sector of a periodic chain.

\section{Errors in forward scattering approximation\label{App:error}}

Here we present a more detailed analysis of the error made in individual steps of the forward-scattering approximation. The vector $\delta w_j$ introduced in Eq.~(\ref{eq:error}) corresponds to the error made in a single FSA iteration. The squared-norm of the error vector can be brought to the following form
\begin{equation}
  \|\delta w_j\|^2 = \langle v_{j-1} | [H_{+},H_{-}] |v_{j-1}\rangle + \beta_j^2 - \beta_{j-1}^2.
  \label{eqn:error_norm}
\end{equation}
From here it is clear that this error is governed by the commutator $[H_{+},H_{-}]$ between forward and backward propagation terms in the Hamiltonian. Explicit evaluation of this commutator gives: 
\begin{multline}  \label{eqn:fsa_commutator}
    [H_{+},H_{-}] = - \sum_j (-)^j P_{j-1}Z_jP_{j+1} \\
    = \hat{D}_{\mathbb{Z}_2} - \frac{L}{2} - \sum_j(-)^jP_{j-1}P_jP_{j+1}, 
\end{multline}
where the operator $\hat{D}_{\mathbb{Z}_2} = \sum_{j \in \text{ odd}} P_j + \sum_{j \in \text{ even}} Q_j$ is diagonal in the basis of product states with eigenvalues giving the Hamming distance of a given product state from  $\ket{\mathbb{Z}_2}$ state. The final term in Eq.~(\ref{eqn:fsa_commutator}) measures the imbalance of ``forward-holes'' and ``backward-holes''. Here we define forward-hole as a pattern ${\circ}{\circ}{\circ}$ centered on a site with odd $j$, where $H_{+}$ could introduce an excitation in the middle. Using the fact that for $j=1,2$ no forward/backward holes exist in the system, and using explicit values of $\beta_{1,2}$ from Eq.~(\ref{eq:betajmain}) one can check that Eq.~(\ref{eqn:error_norm}) indeed gives $ \|\delta w_j\|^2=0$. 

\begin{figure}[tb]
  \centering
  \includegraphics[width=\columnwidth]{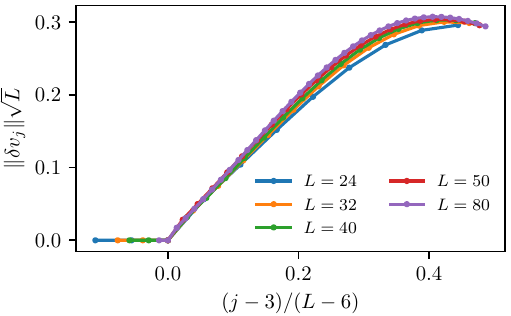}
  \caption{
    Normalized errors $\|\delta v_j\|$ rescaled by a factor of $\sqrt{L}$ for PBC in  in the parity symmetric sector. With this rescaling, the errors for different $L$ are well collapsed, suggesting $\|\delta v_j\|^2 = O(j^2/L^3)$.
  }
  \label{fig:scat_deltav}
\end{figure}

Fig.~\ref{fig:scat_deltav} shows the normalized errors $\|\delta v_j\|$, obtained numerically from the definition in Eq.~(\ref{eq:error}), where we have introduced the error vector $\delta v_j$, defined as $\beta_j \delta v_j = \delta w_j$. This represents the relative error in the Lanczos vector $v_j$. Moreover, in Fig.~\ref{fig:scat_deltav} we have rescaled $\|\delta v_j\|$ by a factor of $\sqrt{L}$. Fig.~\ref{fig:scat_deltav} shows that the error for fixed $j$ decreases with $L$, which is promising for applications of the FSA method to larger systems. However, a more complete error analysis requires an analytical description of how the errors in the individual steps compound to produce final errors in the physical quantities of interest. This is a more challenging question that requires further investigation.

\begin{figure}[t]
  \centering
  \includegraphics[width=\columnwidth]{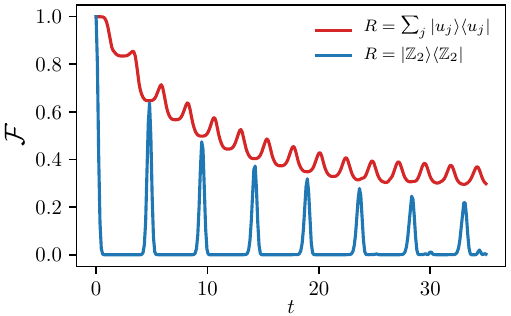}
  \caption{
Red curve shows the  time dependence of the probability to be found within the forward-scattering subspace, measured by Eq.~(\ref{eq:R}). Blue curve shows the fidelity for the N\'eel initial state. Data is for system size $L=32$.
  }
  \label{fig:fid_fsa_sector}
\end{figure}

In Fig.~\ref{fig:fid_fsa_sector} we assess the quality of the forward-scattering approximation for the dynamics. The red curve tracks the probability that the system remains within the forward-scattering subspace over time, starting out in the N\'eel state. This is quantified by calculating the generalization of quantum fidelity defined as
\begin{equation}\label{eq:R}
{\cal F} =  \bra{\mathbb{Z}_2}e^{i H t}Re^{-i H t}\ket{\mathbb{Z}_2},
\end{equation}
where the operator $R = \sum_j |u_j\rangle \langle u_j|$ projects onto the forward-scattering subspace spanned by vectors $|u_j\rangle$. In the same figure, we also show the actual fidelity (return probability) for the N\'eel state, since Eq.~(\ref{eq:R}) reduces to ${\cal F} =  |\bra{\mathbb{Z}_2}e^{-i H t}\ket{\mathbb{Z}_2}|^2,$ when $R=|\mathbb{Z}_2\rangle \langle \mathbb{Z}_2|$. The fact that the generalized fidelity  changes over time indicates that the weight of the wave function contained within the FSA subspace is not invariant under unitary evolution generated by $H$. Equivalently, this implies that the operator $R$ is only an approximate integral of motion. The gap between the revival probability and the subspace probability shows the effect of dephasing within the forward-scattering subspace. Thus, we observe that the leakage of the many-body wave function outside the FSA subspace is the main cause of fidelity decay in a quantum quench. 

\section{Linear recurrence method}\label{sec:loops}

The forward-scattering approximation in the main text was defined by  
the recurrence  in Eq.~(\ref{Eq:lanczosFSA}), which is reproduced here:
\begin{equation}
  \beta_j v_j = H_{+} v_{j-1}\text{.}
  \label{eqn:forwardscattering1}
\end{equation}
Similarly, we must have
\begin{equation}
  \beta_j v_{j-1} = H_- v_j.
  \label{eqn:forwardscattering2}
\end{equation}
Calculating the off-diagonal elements of the Hamiltonian matrix directly from these formulas is, however, inefficient due to an exponential growth of the dimension of the connected subspace. For this reason, in this Appendix we develop a more efficient method. First, in Eq.~(\ref{eq:betaj}), it will be shown that $\beta$ coefficients are determined by the number of loops reaching a given distance from the initial state. Recursive expressions for these loop countings will then be derived for OBC, yielding as main results Eqs.~(\ref{eq:prec_f}), (\ref{eq:prec_l}) and (\ref{eq:Wopen}). Analogous expressions for PBC are given in Eqs.~(\ref{eq:prec_r}), (\ref{eq:prec_m}) and (\ref{eq:Wpbc}). These results allow for an efficient and high-precision implementation of the FSA approximation in large systems on the order of $L\lesssim 100$ sites.
 
\subsection{Off-diagonal matrix elements from loop counting}
 
By repeated application of Eqs.~(\ref{eqn:forwardscattering1})-(\ref{eqn:forwardscattering2}) we obtain
\begin{align}
   \beta_j^2
  &= \langle v_{j-1}| H_{-}H_{+} | v_{j-1}\rangle \notag\\
  &=  \frac{\langle v_0| \left(H_{-}\right)^j\left(H_{+}\right)^j |v_0\rangle}{\prod_{k=1}^{j-1}\beta_k^2} \notag \\
  &=  \frac{\langle v_0| \left(H_{-}\right)^j\left(H_{+}\right)^j |v_0\rangle}{\langle v_0| \left(H_{-}\right)^{j-1}\left(H_{+}\right)^{j-1} |v_0\rangle }\text{.}
\end{align}
From this, we recognize the amplitude
\begin{equation}
  W_{N,j} = \langle v_0| \left(H_{-}\right)^j\left(H_{+}\right)^j |v_0\rangle
  \label{eq:Wj}
\end{equation}
as the number of shortest closed paths reaching a distance $j$ from the initial state. Subscript $N$ indicates the dependence of $W_j$'s on the system size, which we denote by $N$ in this section. In terms of $W$'s, the off-diagonal matrix elements are
\begin{equation}
  \beta_j = \sqrt{\frac{W_{N,j}}{W_{N,j-1}}}\text{.}
  \label{eq:betaj}
\end{equation}
Our goal here is to derive a linear recurrence system for calculating $W_{N,j}$ and our strategy will be to count loops recursively in terms of loops on smaller subsystems. We will first present results for OBC and then generalize to PBC.

\subsection{Loop counting for open chains}

Consider a loop of valid spin flips on an open system of $N$ sites.
This loop can be projected into two subsystems where each flip is assigned to the subsystem in which the spin in flips is located.
We will choose the two subsystems to comprise of the two leftmost spins and the remaining $N-2$ sites of the system.
These subsystem loops are valid loops on the corresponding open systems of $2$ and $N-2$ sites.
The original loop is one of the ways in which the spin flips of the subsystem loops can be interlaced such that the constraints are never violated.
Given a pair of loops on the two subsystems, we only need to know when the left-most spin of the right subsystem and the right-most spin of the left subsystem are flipped, if at all, in order to count ways in which they can be interlaced.
Because the loops discussed are properly shortest loops, these boundary spins are either flipped once moving away from the N\'eel state and once again on the return journey, or not at all.

The most general shortest loop looks like a word
\begin{eqnarray}
   \label{eqn:loopword}
\nonumber   \underbrace{A \cdots A L \underbrace{A \cdots A R \overbrace{A \cdots A}^{c}}_{a}}_{j} \mid \underbrace{\overbrace{\underbrace{A \cdots A}_{b} L A \cdots A}^{d} R A \cdots A}_{j} \\
\end{eqnarray}
where the symbol $A$ represents any spin flip (and each instance is different) of a bulk spin, $L$ represents flipping the leftmost spin and $R$ the rightmost spin.
The vertical line separates the forward and backward steps in the loop.
The order in which the $L$ and $R$ flips appear, if at all, in the forward and backward half-words is not fixed, Eq.~(\ref{eqn:loopword}) represents only one possible ordering.

We will start by considering only the case of open boundaries.
Let $\mathcal{F}$ be the combinatorial class of forward-scattering loops on a system with open boundaries where the left-most spin remains fixed throughout the process.
This class is graded by the size of the system $N$ and the number of forward and backward transitions $j$.
Similarly, let $\mathcal{L}$ be the class where the left-most spin is flipped at some point of the process and is additionally graded by $a$ and $b$.
The index $a$ specifies the number of forward steps which follow the flip of the leftmost spin, $b$ is the number of backward steps preceding the return flip of the leftmost spin.
These classes are defined recursively from the following equations,
\begin{align}
  \mathcal{F}
  &= \begin{tikzpicture}[baseline=+0.25em]
    \node (M) at (0, 0) {${\bullet}{\circ}$};
    \draw (-0.2,0.15) -- (0.2,0.15);
    \node (M) at (0, 0.3) {${\bullet}{\circ}$};
  \end{tikzpicture} \ast \left(\mathcal{F} + \mathcal{L}\right) \label{eq:calF} \\
  \mathcal{L}
  &= \begin{tikzpicture}[baseline=+0.7em]
    \node (M) at (0, 0) {${\bullet}{\circ}$};
    \node (M) at (0, 0.2) {${\circ}{\circ}$};
    \draw (-0.2,0.35) -- (0.2,0.35);
    \node (M) at (0, 0.5) {${\circ}{\circ}$};
    \node (M) at (0, 0.7) {${\bullet}{\circ}$};
  \end{tikzpicture} \ast \left(\mathcal{F} + \mathcal{L}\right)
  + \begin{tikzpicture}[baseline=+1.2em]
    \node (M) at (0, 0) {${\bullet}{\circ}$};
    \node (M) at (0, 0.2) {${\circ}{\circ}$};
    \node (M) at (0, 0.4) {${\circ}{\bullet}$};
    \draw (-0.2,0.55) -- (0.2,0.55);
    \node (M) at (0, 0.7) {${\circ}{\bullet}$};
    \node (M) at (0, 0.9) {${\circ}{\circ}$};
    \node (M) at (0, 1.1) {${\bullet}{\circ}$};
  \end{tikzpicture} \ast \mathcal{L} \label{eq:calL}
\end{align}
where the $\ast$ operation glues a two-site system onto the left, and for each pair of elements in the classes appearing on the left and right hand sides produces all the interlacings that satisfy the Fibonacci constraint.
The vertical direction in the loop diagrams are successive steps in the loop and the horizontal line separates the forward steps from the backward steps.

These equations can be made explicit by introducing the counting sequences for the classes.
Let $f_{N,j}$ be the number of shortest loops on the open chain of $N$ sites reaching a Hamming distance of $j$ from the N\'eel state where the leftmost spin is invariant, and let $l_{N,j}^{a,b}$ count those loops where the leftmost spin is flipped.
These are the counting sequences for $\mathcal{F}$ and $\mathcal{L}$, respectively.
The previous equations (\ref{eq:calF})-(\ref{eq:calL}) then become
\begin{align}
  f_{N,j} &= f_{N-2,j} + \sum_{a',b'} l_{N-2,j}^{a',b'},
  \label{eq:prec_f} \\
  l_{N,j}^{a,b} &= f_{N,j-1} + \sum_{a',b'} T^{a,a'} l_{N-2,j-2}^{a',b'} T^{b,b'},
  \label{eq:prec_l}
\end{align}
where we have introduced $T^{a,a'} = \min(a,a'+1)$.
Eq.~(\ref{eq:prec_f}) captures the idea that we may glue two additional sites and the loop of doing nothing onto any loop on the reduced system.
Eq.~(\ref{eq:prec_l}) captures the idea that the loop that flips only the leftmost spin may be glued to any loop on the reduced system, but when the interior site is excited in the left-subsystem then the leftmost site right-subsystem must first have its excitation removed.
Finally, the class of all loops in $\mathcal{F} + \mathcal{L}$ can be counted by
\begin{equation}
  W_{N,j} = f_{N+2,j}\text{,}
  \label{eq:Wopen}
\end{equation}
because for every loop on $N$ sites we may glue the trivial loop on two sites onto its left boundary to get a distinct loop on $N+2$ sites, and for every loop on $N+2$ sites we may cut off the leftmost two sites to get a distinct loop on $N$ sites.

Let us illustrate how the recurrence works on an example with $N=4$ site open chain. Directly from Eq.~(\ref{eq:Wj}), it is easy to show that the number of loops for different $j$ sectors is given by $W_{4,1}=2$, $W_{4,2}=5$, $W_{4,3}=13$, and $W_{4,4}=25$.  Using the recurrence, we can obtain these values starting from a smaller $N=2$ site chain. In that case, the admissible $j$ are given by 0, 1 and 2, and the only non-zero coefficients are $f_{2,0}=l_{2,1}^{0,0}=l_{2,2}^{1,1}=0$. Then, applying Eqs.~(\ref{eq:Wopen}), (\ref{eq:prec_f}), and (\ref{eq:prec_l}), we have
\begin{eqnarray}
  W_{4,1} &=& f_{6,1} = f_{4,1} + \sum_{a,b=0} l_{4,1}^{a,b}, \notag \\
  f_{4,1} &=& \sum_{a',b'}l_{2,1}^{a',b'} = 1, \notag \\
  l_{4,1}^{a,b} &=& f_{4,0} = 1, 
\end{eqnarray}
thus we get $W_{4,1} = 1 + 1 = 2$.  Analogous calculation gives $W_{4,2}=5$. For $W_{4,3}$ we need to include the $T$ tensors. We have
\begin{eqnarray}
W_{4,3} &=& f_{6,3} = f_{4,3} + \sum_{a,b=0}^2 l_{4,3}^{a,b}, \notag \\
f_{4,3} &=& 0, \notag \\
l_{4,3}^{a,b} &=& f_{4,2} + \sum_{a',b'} T^{a,a'} T^{b,b'} l_{2,1}^{a',b'} = 1 + T^{a,0}T^{b,0}.\qquad
\end{eqnarray}
Therefore, we obtain 
\begin{equation}
W_{4,3} = \smashoperator[lr]{\sum_{a,b=0}^2} \left(1 + \min(a,1)\min(b,1)\right) = 9 + 4 = 13,
\end{equation}
as anticipated. Repeating this procedure and using Eq.~(\ref{eq:betaj}), we can obtain the set of $N+1$ coefficients $\beta_j$ that form the tridiagonal matrix in the forward-scattering approximation.

\begin{figure}[tb]
  \centering
  \includegraphics[width=\columnwidth]{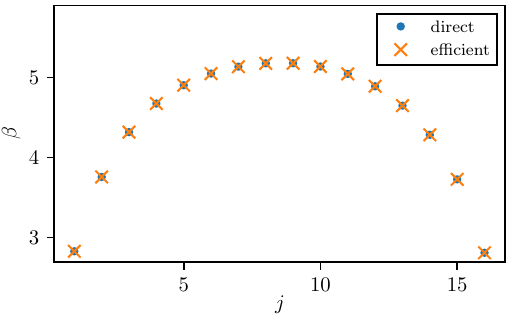}
  \caption{
    A demonstration of exact agreement between direct computation of $\beta$ coefficients from Eq.~(\ref{eqn:forwardscattering1}) and computation from the efficient recurrence, Eq.~(\ref{eq:Wopen}).
    Data is for $N=16$ site chain with open boundary conditions.
  }
  \label{fig:compare_beta_open}
\end{figure}

The $\beta$ calculated by the linear recurrence system, Eqs.~(\ref{eq:prec_f}) and (\ref{eq:prec_l}),  are compared against direct computation with Eq.~(\ref{eqn:forwardscattering1}) in Fig.~\ref{fig:compare_beta_open}. The two methods indeed agree to machine precision for the given system size, but the recurrence method provides much higher accuracy in larger system sizes. Computation time for calculating $\beta_{N,N/2}$ for an open chain of $N$ sites is found to scale roughly as $\propto N^5$.

\subsection{Loop counting on periodic chains}

For periodic boundaries we must keep track of what happens at the subsystem right boundary, to retain the information required to ensure the constraints are not violated when the left and right sides are glued together.
We reinterpret $\mathcal{F}$ and $\mathcal{L}$ with the additional requirement that for the loops in theses classes the rightmost site is at no point excited.
This gives us two new classes, $\mathcal{R}$ and $\mathcal{M}$, which are the analogs of $\mathcal{F}$ and $\mathcal{L}$, except that the rightmost site is now excited at some point of the process.
These classes are graded by indices $c$ and $d$ which mark when the the right boundary site is excited and unexcited.
The new classes satisfy analogous equations to the previous classes
\begin{align}
  \mathcal{R}
  &= \begin{tikzpicture}[baseline=+0.3em]
    \node (M) at (0, 0) {${\bullet}{\circ}$};
    \draw (-0.2,0.15) -- (0.2,0.15);
    \node (M) at (0, 0.3) {${\bullet}{\circ}$};
  \end{tikzpicture} \ast \left(\mathcal{R} + \mathcal{M}\right) \\
  \mathcal{M}
  &= \begin{tikzpicture}[baseline=+0.6em]
    \node (M) at (0, 0) {${\bullet}{\circ}$};
    \node (M) at (0, 0.2) {${\circ}{\circ}$};
    \draw (-0.2,0.35) -- (0.2,0.35);
    \node (M) at (0, 0.5) {${\circ}{\circ}$};
    \node (M) at (0, 0.7) {${\bullet}{\circ}$};
  \end{tikzpicture} \ast \left(\mathcal{R} + \mathcal{M}\right)
  + \begin{tikzpicture}[baseline=+0.9em]
    \node (M) at (0, 0) {${\bullet}{\circ}$};
    \node (M) at (0, 0.2) {${\circ}{\circ}$};
    \node (M) at (0, 0.4) {${\circ}{\bullet}$};
    \draw (-0.2,0.55) -- (0.2,0.55);
    \node (M) at (0, 0.7) {${\circ}{\bullet}$};
    \node (M) at (0, 0.9) {${\circ}{\circ}$};
    \node (M) at (0, 1.1) {${\bullet}{\circ}$};
  \end{tikzpicture} \ast \mathcal{M}\text{.}
\end{align}
This is because the gluing process never changes whether the rightmost site is excited or left invariant.
 
Keeping track of how all these indices are changed as loops are interlaced during gluing results in the following set of equations:
\begin{align}
  \label{eq:prec_r}
  r_{N,j}^{c,d}
  &= r_{N-2,j}^{c,d} + \sum_{a',b'} m_{N-2,j}^{a',b',c,d}\text{,} \\
  m_{N,j}^{a,b,c,d}
  &= r_{N,j-1}^{c-\delta(c \ge a),\,d-\delta(d \ge b)} \nonumber \\
  &\;\; + \sum_{a',b',c',d'}T^{c,c',a,k} m_{N-2,j-2}^{a',b',c',d'} T^{d,d',b,m}\text{,}
  \label{eq:prec_m}
\end{align}
where
\begin{equation}
  T^{a,c,a',c'} = \delta_{a \neq c} \smashoperator[lr]{\sum_{k=0}^{\min(a',a-1)}} \delta_{k \neq c}\,\delta_{c',\, c - \delta(c \ge a) - \delta(c \ge k)}.
\end{equation}
Finally, the total number of loops is found by demanding compatibility between the left and right boundaries
\begin{equation}\label{eq:Wpbc}
  W_{N,j} = f_{N,j} + \sum_{a,b} l_{N,j}^{a,b} + \sum_{a > c,\,b > d} m_{N,j}^{a,b,c,d}\text{.}
\end{equation}

\end{document}